\documentclass{elsart}
\usepackage{amssymb}
\usepackage{graphicx}
\usepackage[colorlinks=true, linkcolor = blue urlcolor = blue, citecolor = blue]{hyperref}
 \def\ket{\!>\,}

\usepackage{longtable}
\usepackage{multirow}
\usepackage{longtable}
\usepackage[english]{babel}
\usepackage{amssymb}
\usepackage{amsmath}
\usepackage{graphicx}
\usepackage{subfigure}
\usepackage{mathptmx}

\usepackage{epsfig}
\usepackage{epstopdf}
\usepackage{color,amsmath,amssymb,epsfig,graphicx}

\usepackage{graphicx}
\usepackage{longtable}

\usepackage{mathptmx, courier, pifont}
\usepackage[scaled=0.92]{helvet}
\usepackage[T1]{fontenc}
\usepackage{textcomp}

\ExplSyntaxOn \cs_gset:Npn \__first_footerline: { \group_begin: \small \sffamily \__short_authors: \group_end: } \ExplSyntaxOff
\begin{document}

\begin{frontmatter}
\title{Microscopic investigation of wobbling motion in atomic nuclei using the triaxial projected shell model approach}

\author{J.A. Sheikh$^{1,2}$, S. Jehangir$^{2}$ and G. H. Bhat$^{3}$}
\address{$^1$ Department of Physics, University of Kashmir, Srinagar,
190 006, India \\
$^2$ Department of Physics, Islamic University of Science and Technology, Kashmir, India\\
$^3$ Department of Physics, Govt. Degree College, Shopian, India
 }

\begin{abstract}
A systematic investigation of the wobbling band structures observed in  odd-mass nuclei of $^{161,163,165,167}$Lu, $^{167}$Ta  $^{131}$Cs, $^{135}$Pr, $^{151}$Eu, $^{183}$Au, $^{133}$Ba, $^{105}$Pd, $^{133}$La, $^{187}$Au and $^{127}$Xe is performed using the triaxial projected shell model (TPSM) approach. It is demonstrated that all the studied band structures have transverse wobbling mode, except for $^{133}$La, $^{187}$Au (negative parity), $^{183}$Au (positive parity) and $^{127}$Xe nuclei where the wobbling frequency increases with spin, indicating
 that the collective motion has a longitudinal character. To elucidate further the wobbling nature of the band structures, electromagnetic transition probabilities have been evaluated
and it is observed that  inter-band transitions are dominated by $E2$ rather than $M1$ as expected for a typical signature partner band. It is shown that TPSM approach provides a reasonable description of all the measured properties of the studied nuclei. 
\end{abstract}
\end{frontmatter}
\section{Introduction}

Atomic nuclei depict a rich diversity of shapes and structures, primarily driven by the quantum mechanical
shell effects and to elucidate these features microscopically is one of main research themes in nuclear physics \cite{SF2015}.
The information regarding the nuclear shape is inferred through a comparison of the measured quantities with the predictions
of a theoretical model that assumes a geometrical shape for the atomic nucleus in the intrinsic frame of reference. The unified model
of Bohr and Mottelson has played a key role in our understanding of nuclear shapes \cite{BM}. In this model, the 
shape is assumed to have dominant quadrupole degree of freedom and is parametrized in terms of axial and
non-axial deformation parameters of $\beta$ and $\gamma$. Most of the deformed nuclei have prolate shapes with $\gamma=0^\circ$, and there
are also some nuclei with oblate spheroidal shape having $\gamma=60^\circ$. These nuclei are referred to as axial and have
the projection of the angular momentum along the symmetry axis, designated by ``K'', a conserved quantum number. There are selection rules based on this conserved
quantity, for instance, Alaga rules and nuclei in many regions of the periodic table are known to follow the selection rules
based on the K-quantum number \cite{BM}.

Nevertheless, there are also a few regions in nuclear chart for which axial symmetry is predicted to be broken with $0 < \gamma^\circ < 60$.
The regions include $A\sim 160$ \cite{WM1,WM44,WM45,WM2,HA05,Aberg1990,IR1989,TB1989} and 
$A\sim 130$ \cite{WM54,WM47,WM48,WM6,WM12}. These nuclei, referred to as triaxial systems, have unequal
moments of inertia along the three principle axes (short axis {\it s}, medium axis {\it m} and long axis {\it l}) 
and the rotational motion can lead to very special band structures corresponding to chiral and wobbling phenomena.
The chiral symmetry is expected for nuclei where total angular momentum have components along all the three mutually perpendicular axes
with the collective rotor angular momentum directed along the medium axis, which has the largest moment of inertia within the
irrotational model and $0 \le \gamma \le 60$. The
angular momentum vectors of the valence particles align their angular momentum vectors along the short and long axis for particle-
and hole-orbits, respectively. This arrangement of the three mutually perpendicular angular momentum vectors form two coordinate systems
with opposite chirality, namely, with left and right handedness, corresponding to the left and right hand orientations of the three vectors.
The operator transforming these two systems known as chiral operator ($\hat{\chi} = \hat{\mathcal T}\hat{ \mathcal R}(\pi)$), first performs a rotation
through $180^\circ$ and then the time reversal operation, changing the directions of the angular momenta. The spontaneous
breaking of chiral symmetry occurs in the intrinsic or body-fixed reference frame and in the laboratory frame  with the restoration of
this  broken symmetry, it manifests in the appearance of chiral doublet bands, i.e., a pair $\Delta I = 1\hbar$ bands with the same parity.

Another observation that is only possible for triaxial systems is the wobbling motion and occurs when the rotation about the
intermediate or medium axis having largest moment of inertia precesses about the fixed angular momentum vector because of the perturbations
from the rotation of short and long axis. The wobbling motion, the classical analog of which is the spinning motion
of an asymmetric top \cite{IH01}, is an excitation mode unique to a
triaxial body. The rotation about the medium axis with the largest moment of inertia has the
lowest energy and gives rise to the yrast states. The excited states near the yrast line are generated by transferring some
angular momentum from the {\it m}-axis to {\it s}- and {\it l}-axes. It was shown by Bohr and Mottelson that for large angular momentum  
$I >>1$, a family of rotational bands are obtained which are characterized by the wobbling quantum number $(n_{\omega})$. Considering the
$D_2$ symmetry, one obtains the selection rule $:(-1)^{n_{\omega}} = (-1)^{I}$ \cite{BM} for even-even systems. Therefore, the band structures with
even $n_{\omega}$ ($n_{\omega}=0,2,4..$) contain only even-spin states, and with odd $n_{\omega}$ ($n_{\omega}=1,3,...$) have only
odd-spin members. The key characteristic 
feature of the wobbling bands is that the interband I->(I-1) transitions are predominantly E2 as compared with the transitions between the two signatures in normal rotational bands which are predominantly M1. 

The wobbling excitation mode was first identified in $^{163}$Lu \cite{WM1} and subsequently was observed in $^{161,165,167}$Lu
\cite{WM59,WM44,WM45} and $^{165}$Tm isotopes \cite{WM46}. Recently, the wobbling mode has also been observed in other mass regions \cite{HA05,WM54,WM47,WM48,WM6,WM12}.
The purpose of the present work is to perform a systematic investigation of the wobbling band structures observed in odd-mass
isotopes using the microscopic approach of the triaxial projected shell model (TPSM)  \cite{TPSM1999}. This model is now well established
to provide a unified description of the properties of deformed and transitional nuclei \cite{TPSM8,TPSM2014,TPSM5,TPSM4,TPSM3,TPSM7,TPSM2,TPSM6}. The TPSM approach has been
employed to perform a systematic investigation of the chiral band structures observed in various mass regions of the periodic
table \cite{WM39}. In a more recent work, a systematic analysis has been carried out for the $\gamma$ bands observed in
deformed nuclei and it has been demonstrated that staggering of the energies and $B(E2)$ transition probabilities
of the $\gamma$ bands can provide important information on the nature of the collective motion in atomic nuclei \cite{TPSM2024}.

The present chapter is organized in the following manner. In the next section, we provide some basic elements of the wobbling
motion as discussed in the original work of Bohr and Mottelson for even-even systems, and generalized to
odd-mass systems by Hamamoto \cite{IH01} and Frauendorf and D\"{o}nau \cite{WM17}. In section 12.3, a survey of experimental
results on the wobbling motion is provided. In section 12.4, the microscopic TPSM approach used in the present work is briefly
discussed. In section 12.5, the results of the TPSM are presented and discussed
in light of the available experimental measurements. In section 12.6, the present investigation on wobbling mode is summarized
with some future perspectives.
\section{Basic concepts of wobbling mode}

Bohr and Mottelson in 1975 showed that wobbling mode in atomic nuclei for large angular-momentum, $I >> 1$, will give rise to a family
of band structures characterized by the phonon quantum number \cite{BM}. The
properties of the wobbling bands were obtained using the triaxial particle-rotor model (TPRM) and in the present section,
we shall  follow the original work of Bohr and Mottelson \cite{BM}.

Considering that the triaxial deformed field is time-reversal invariant, the leading order rotor Hamiltonian for an even-even system
can be written as 

\begin{equation}
{\widehat H}_{rot} = \sum_{\kappa=1}^{3}~A_{\kappa}~{\hat I}^{~2}_\kappa~~~~  ,
\end{equation}
where the coefficients $A_{\kappa}$ are inverse of the moments of inertia (${\mathcal{I}}_\kappa$) and are given by
\begin{equation}
A_\kappa = \frac {\hbar^2} { 2 {\mathcal{I}}_\kappa }~~~~.
\end{equation}
 The axes are labeled such that $A_1 < A_2 < A_3$ or accordingly $\mathcal{I}_1 >\mathcal{I}_2 >\mathcal{I}_3$ and the yrast states  have $ |I_{1}|  \sim I$.

The wobbling excitations are small amplitude oscillations of the angular momentum vector about the axis having the largest moment of inertia. The energy values are given by a harmonic spectrum of wobbling quanta \cite{BM}
\begin{equation}
E(n_\omega, I) = A_1I(I+1) + (n_\omega
+ \frac{1}{2})~\hbar\omega ,
\end{equation}
where the quantum number $n_\omega$ gives the number of wobbling quanta and $\hbar\omega$ is the wobbling energy defined as the energy associated with wobbling excitation and is given by 
\begin{equation}
\hbar\omega = 2I((A_2 - A_1)( A_3 - A_1))^{1/2}
\end{equation}
The quadrupole transition probabilities are given by \cite{BM}
\begin{eqnarray}
& B(E2;n_{\omega}I\rightarrow n_{\omega},I\pm 2) \approx\frac{5}{16\pi}e^2Q_2^2 \label{be21}\\
&B(E2;n_{\omega}I\rightarrow n_{\omega}-1,I-1) =\frac{5}{16\pi}e^2\frac{n_{\omega}}{I}(\sqrt{3}Q_0x-\sqrt{2}Q_2y)^2\label{be22}\\
  &B(E2;n_{\omega}I\rightarrow n_{\omega}+1,I-1) =\frac{5}{16\pi}e^2(\frac{n_{\omega}+1}{I})(\sqrt{3}Q_0y-\sqrt{2}Q_2x)^2\label{be23}
\end{eqnarray}
where, $Q_0$ and $Q_2$ are  intrinsic quadrupole moments.

The excitations in wobling mode are based on the quantum phonon
model and have very special characteristics. In particular, the transitions from the $n_\omega=2$ to the $n_\omega=1$ should be a factor of two
stronger as compared to the transitions from  $n_\omega=1$ to  $n_\omega=0$. On the other hand, the direct transition from  $n_\omega=2$ to
$n_\omega=0$ band are forbidden.

It was shown in the Refs. \cite{IH01} and \cite{WM17} that the presence of an odd-particle will modify the properties of wobbling motion in odd-mass systems. In particular, if the odd-particle occupies a low-$K$ orbital of a high-$j$ intruder orbital, the wobbling frequency will decrease \cite{WM17}
with angular-momentum, $I$ and is referred to as the transverse wobbling (TW). This is in comparison to the longitudnal wobbling (LW),
originally proposed by Bohr and Mottelson, for which the wobbling frequency increases with I. For odd-mass systems, the LW mode occurs if the odd-particle
occupies a high-$K$ orbital.

\section{Survey of experimental data on wobbling mode }

Although wobbling motion was originally predicted \cite{BM} for an even-even nucleus, but it was first  experimentally established
in the odd-proton nucleus $^{163}$Lu  \cite{WM1}.  For this
system, several triaxial strongly deformed (TSD)
band structures were known to exist \cite{WM41}, however, the structure of
these bands remained obscure as the transitions between the
bands were not measured. In the seminal work \cite{WM1},  nine interconnecting transitions
were measured and it was possible to
establish the structure of the two TSD bands, referred to as TSD1 and TSD2. It was demonstrated that these two bands have similar intrinsic structure
with alignments $(i_x)$ and moments of inertia (${J}^{(2)}$) being almost identical. In the normal cranking picture \cite{WM56}, these bands
are signature partner bands with TSD1
being the favoured partner having $\alpha=1/2$ and TSD2 having $\alpha=-1/2$ with $i_x$ of the later being different from the former as in the
unfavoured branch, the aligned angular-momentum of the particle is tilted with respect to the rotational axis. In the case of the favoured signature
band, the angular-momentum  vector of the particle is aligned along the rotational axis. Further, in the cranking model
the unfavored
signature band is expected at a higher excitation energy greater than $1~$MeV and in the experimental data, its excitation energy relative to
TSD1 band is less than $0.3~$MeV. It has been shown using the particle-rotor model description that the observed features in $^{163}$Lu could be explained
by using three moments of inertia that correspond to the wobbling motion of a triaxial nucleus \cite{WM21,WM57}.


Electromagnetic properties provide a stringent test on the nature of the excitation mechanism observed in atomic nuclei \cite{BM}. In the cranking
picture, $B(M1)$ transitions between $\alpha=-1/2$ and $+1/2$ signature partner bands are enhanced, whereas for the wobbling motion these
$I \rightarrow (I-1)$ transitions are dominated by $B(E2)$. The angular-correlation, angular-distribution and polarization
data \cite{WM1} were found to be consistent with
$M1/E2$ multipolarity for the connecting transitions with the mixing ratio ($\delta$) of $(90.6 \pm 1.3)\%$ for the $E2$ and $(9.4 \pm 1.3)\%$ for
the $M1$ transition. It was also noted that the phase of the zigzag pattern observed for $B(E2)_{out}/B(E2)_{in}$ and $B(M1)_{out}/B(E2)_{in}$ was
opposite to that calculated in the cranking model and was consistent with the phase obtained for the wobbling mode in particle-rotor model
calculations \cite{WM1}. Moreover, in the quantal phonon model of the wobbling mode, which is valid for $I >>1 \hbar$, the inter-band transitions follow
special features with $B(E2)$ of the inter-band  $I \rightarrow (I-1)$ transitions competing with those of the in-band $I \rightarrow (I-2)$ transitions. Further, the inter-band
$B(E2)$ strength from the $n_{\omega}=2$ to the $n_{\omega}=1$ band should be a factor of two larger than the $B(E2)$ strength from
the $n_{\omega}=1$ to the $n_{\omega}=0$ band \cite{BM,WM21,WM57}. These expected properties of the phonon wobbling mode have been experimentally verified
for $^{163}$Lu \cite{WM42}.

Candidate wobbling bands have also been reported in other neighbouring odd-A lutetium isotopes of $^{161,165,167}$Lu
\cite{WM44,WM45,WM2,HA05}, although
the angular correlation
and linear polarization measurements were not possible to confirm the $B(E2)$ character of the inter-band transitions as has been done for
the $^{163}$Lu isotope.
Nevertheless, the similarity of the decay pattern of the TSD band structures of the three Lu isotopes with  $^{163}$Lu and large values
obtained for
the $B(E2)_{out}/B(E2)_{in}$ transitions, the structures have been designated as wobbling bands. The search for wobbling in other nuclides
in this mass region failed to establish any more candidates, except for $^{167}$Ta \cite{WM46}. It was demonstrated
using the tilted axis cranking approach that the density of states for Lu isotopes is quite low for the TSD minimum and the wobbling
bands could be identified. In other neighbouring nuclides, the density of sates is quite high and particle-hole excitation modes are more
favourable \cite{WM58}.
For $^{167}$Ta, the similarity of the linking transitions between the TSD2 and TSD1 bands with the corresponding transitions in $^{163,165,167}$Lu
isotopes, the two TSD bands observed have been grouped as wobbling structures \cite{WM46}. We would like to
mention that TSD1 band in  $^{167}$Ta depicts a bandcrossing feature with
${ J}^{(2)}$  showing a peak at about $\hbar \omega = 0.35~$ MeV. This bandcrossing
feature is also observed in $^{167}$Lu, but not in other Lu isotopes.

In all the above systems, the wobbling bands have been established for the TSD minimum configuration with axial deformation
value of $\sim 0.40$. 
The wobbling mode for a normal deformed nuclide was reported for the first time in $^{135}$Pr with $\epsilon \sim 0.16$ \cite{WM47}.
Wobbling and signature partner (SP) band structures were delineated from the analysis of the angular distribution and polarization data.
The mixing ratios
determined from these measurements show large $B(E2)$ mixing for the transitions from $n_{\omega}=1$ wobbling band to the yrast structure and for the SP band it is
dominated by $B(M1)$. The evidence for the existence
of a two-phonon wobbling band was provided in a subsequent work \cite{WM48}. It needs to be
added that new measurements have been performed and from the analysis of this new data, it has been shown that mixing ratio,
$|\delta| < 1$, contradicting the wobbling interpretation of the observed bands in  $^{135}$Pr
\cite{WM49}.

It was predicted \cite{WM17} using the particle-rotor model that for the LW wobbling mode, the
wobbling frequency initially will increase upto some critical angular-momentum and then will begin decreasing. However, in all the
TW cases identified before 2020, wobbling frequency  decreased with spin. It was shown in 2020 \cite{WM51}
that for one of the two TW wobbling bands identified in $^{183}$Au, the frequency decreases initially and then after some spin values it increases.
For the heavier $^{187}$Au, a pair of longitudinal wobbling structures have been observed with the SP band \cite{WM52}.

The appearance of wobbling bands have also been reported in $^{127}$Xe and $^{133}$La through angular distributions and linear polarization measurements.
The bands observed have been classified as $n_{\omega} =0, 1, 2$ and SP bands
\cite{WM54,WM53}. In both these nuclei, the wobbling frequency decreases
with spin and, therefore, has transverse character with angular-momentum of the particle along the {\it s}-axis. In $^{133}$Ba, longitudinal wobbling
mode has been proposed as for $^{135}$Pr with wobbling frequency decreasing as a function of spin \cite{WM55}. 

\section{Triaxial projected shell model approach}

Triaxial projected shell model approach is conceptually similar to the  spherical shell model (SSM) and differing only in the way the basis space is
chosen. TPSM employs the deformed intrinsic states of the triaxial Nilsson potential, which incorporate essential long-range 
correlations, as the basis configuration \cite{TPSM1}. The truncation of the many-body basis in TPSM is very efficient as not only
the numerical effort is drastically reduced, but also makes physical interpretation more transparent.
The recently generalized TPSM approach with the inclusion of higher-order quasiparticle states has emerged as a powerful tool for exploring
the triaxial characteristics of atomic nuclei as computational resources involved are quite modest and it is feasible
to systematically investigate a broad range of atomic nuclei.  
As a matter of fact, several systematic investigations have been performed for chiral and $\gamma$  vibrational band structures
observed in triaxial nuclei \cite{TPSM8,TPSM2014,TPSM5,TPSM4,TPSM3,TPSM7,TPSM2,TPSM6}. The model space in the TPSM approach is spanned by multiquasiparticle basis states from different oscillator shells \cite{TPSM10,TPSM9}. This allows to investigate high-spin band structures in well-deformed and transitional nuclei all across the nuclear chart.

The flow chart of the TPSM calculations can primarily be divided into three stages. In the first stage, the deformed basis are
constructed by solving the triaxial Nilsson potential with a realistic set of quadrupole deformation parameters of
$\epsilon$ and $\epsilon^{\prime}$ for a given system under consideration. These deformation values lead to an accurate Fermi surface
and it allows one to choose an optimum set of the basis states around the Fermi surface for a realistic description of the system. The standard
Bardeen–Cooper– Schriefer (BCS) procedure is then carried out to include the pairing correlations with the parameters given in Table \ref{Tab:Tcr}.

The rotational symmetry is not conserved by the intrinsic states generated from the deformed Nilsson calculations and in the second stage this symmetry is restored through three - dimensional angular-momentum projection technique \cite{TPSM11,TPSM12}. In this technique, good angular-momentum basis states are projected out from the Nilsson + BCS states using the explicit three- dimensional  angular-momentum projection operator
''$P^I_{MK}$''  given by
  \cite{TPSM13} $:$ 
\begin{equation}
\hat P ^{I}_{MK}= \frac{2I+1}{8\pi^2}\int d\Omega\, D^{I}_{MK}
(\Omega)\,\hat R(\Omega),
\label{proj}
\end{equation}
with the rotation operator 
\begin{equation}
\hat R(\Omega)= e^{-i\alpha \hat J_z}e^{-i\beta \hat J_y}
e^{-i\gamma \hat J_z}.\label{rotop}
\end{equation}
Here, $''\Omega''$ represents the set of Euler angles 
($\alpha, \gamma = [0,2\pi],\, \beta= [0, \pi]$) and  
$\hat{J}^{,}s$ are the angular-momentum operators. The
angular-momentum projection operator in Eq.~(\ref{proj}) apart from 
projecting the good angular-momentum, also projects states 
with good $K$-values.
The projected multi-quasiparticle basis states for different systems  are given 
for odd-proton nuclei as $:$
\begin{eqnarray}\nonumber
  &&\{\hat P^I_{MK}~a^\dagger_{\pi_1}|\Phi\rangle;  \hat P^I_{MK}~a^\dagger_{\pi_1}a^\dagger_{\nu_1} a^\dagger_{\nu_2} |\Phi\rangle ; \hat P^I_{MK}~a^\dagger_{\pi_1} a^\dagger_{\pi_2} a^\dagger_{\pi_3}
 |\Phi\rangle; \\
&& \hat P^I_{MK}~a^\dagger_{\pi_1} a^\dagger_{\pi_2} a^\dagger_{\pi_3}
a^\dagger_{\nu_1} a^\dagger_{\nu_2}|\Phi\rangle  \} \label{basis2}~~~~~~~~~~~~,
\end{eqnarray}    
and for odd-neutron nuclei $:$
\begin{eqnarray}\nonumber
  &&\{\hat P^I_{MK}~a^\dagger_{\nu_1}|\Phi\rangle;  \hat P^I_{MK}~a^\dagger_{\nu_1}a^\dagger_{\pi_1} a^\dagger_{\pi_2} |\Phi\rangle ; \hat P^I_{MK}~a^\dagger_{\nu_1} a^\dagger_{\nu_2} a^\dagger_{\nu_3}
 |\Phi\rangle;\\
&& \hat P^I_{MK}~a^\dagger_{\nu_1} a^\dagger_{\nu_2} a^\dagger_{\nu_3}
a^\dagger_{\pi_1} a^\dagger_{\pi_2}|\Phi\rangle  \}\label{basis3}~~~~~~~,
\end{eqnarray} 
where $ |\Phi\rangle$ in above equations represent the triaxial quasiparticle vacuum state and $a^\dagger_{\nu_i}$, $a^\dagger_{\pi_i}$ are quasiparticle creation operators, with the index ${\nu_i}({\pi_i})$, denoting the neutron (proton) quantum numbers and running over the selected single-quasiparticle states.

In the third and final stage,  projected basis states given by Eqns.~ (\ref{basis2}) and (\ref{basis3})
are then used to diagonalize the shell model Hamiltonian. A two-body Hamiltonian in terms of separable forces is adopted which consists of the
modified harmonic oscillator single-particle Hamiltonian and a  
residual two-body interaction comprising of quadrupole-quadrupole, monopole pairing  and quadrupole pairing terms. 
These terms represent specific correlations which are considered to be essential to describe the low-energy nuclear phenomena \cite{TPSM14}. The Hamiltonian has the following form $:$
\begin{eqnarray}
\hat H =  \hat H_0 -   {1 \over 2} \chi \sum_\mu \hat Q^\dagger_\mu
\hat Q^{}_\mu - G_M \hat P^\dagger \hat P - G_Q \sum_\mu \hat
P^\dagger_\mu\hat P^{}_\mu . \label{hamham}
\end{eqnarray}
In the above equation, $\hat H_0$ is the spherical single - particle Hamiltonian containing the proper spin–orbit force for correct shell closures \cite{TPSM15}. 
The QQ-force strength, $\chi$, in Eq. (\ref{hamham}) is related to
the quadrupole deformation $\epsilon$ as a result of the
self-consistent Hartree-Fock-Bogoliubov condition and the relation is given by
\cite{TPSM1}:
\begin{equation}
\chi_{\tau\tau'} =
{{{2\over3}\epsilon\hbar\omega_\tau\hbar\omega_{\tau'}}\over
{\hbar\omega_n\left<\hat Q_0\right>_n+\hbar\omega_p\left<\hat
Q_0\right>_p}},\label{chi}
\end{equation}
where $\omega_\tau = \omega_0 a_\tau$, with $\hbar\omega_0=41.4678
A^{-{1\over 3}}$ MeV, and the isospin-dependence factor $a_\tau$ is
defined as
\begin{equation}
a_\tau = \left[ 1 \pm {{N-Z}\over A}\right]^{1\over 3},\nonumber
\end{equation}
with $+$ $(-)$ for $\tau =$ neutron (proton). The harmonic
oscillation parameter is given by $b^2_\tau=b^2_0/a_\tau$ with
$b^2_0=\hbar/{(m\omega_0)}=A^{1\over 3}$ fm$^2$. The monopole pairing strength $G_M$ (in MeV)
is of the standard form
\begin{eqnarray}
G_M = {{G_1 \mp G_2{{N-Z}\over A}}\over A}, 
 \label{pairing}
\end{eqnarray}
where the minus (plus) sign applies to neutrons (protons).The values of  $G_1$ and $G_2$ are chosen such that the 
calculated gap parameters reproduce the experimental mass differences. The quadrupole pairing strength $G_Q$ is assumed to be 
proportional to $G_M$, the proportionality constant being fixed as usual to be 0.16. The single-particle space for the TPSM 
usually includes three  major harmonic oscillator shells (N) each for neutrons and protons in a calculation for deformed heavy nuclei. This large size of single-particle space accommodates 
sufficiently large number of active nucleons as well as all the important orbits and ensures the microscopic description of the collective motion.  For different mass regions the oscillator shells employed are:  N = 4, 5, 6 for neutrons and N = 3, 4, 5, for protons  for A =160 and 180 region; N = 3, 4, 5 for neutrons and N = 3, 4, 5, for protons  for A =130 region; N = 3, 4, 5 for neutrons and N = 2, 3, 4 for protons  for A =100 and 110 regions. 
The shell
model Hamiltonian,  Eq.~(\ref{hamham})  is diagonalized in the angular-momentum projected basis states,
Eqns.~ (\ref{basis2}) and (\ref{basis3}), by following the  
 Hill-Wheeler prescription \cite{TPSM1}. The generalized eigen-value
equation is given by 
\begin{equation}
 \sum_{\kappa^{'}K^{'}}\{\mathcal{H}_{\kappa K \kappa^{'}K^{'}}^{I}-E\mathcal{N}_{\kappa K 
\kappa^{'}K^{'}}^{I}\}f^{\sigma I}_{\kappa'K'}=0, \label{a15}
\end{equation}
 where the Hamiltonian and norm kernels are given by
 \begin{eqnarray*}
 && \mathcal{H}_{\kappa K \kappa^{'}K^{'}}^{I} = \langle \Phi_{\kappa}|\hat H\hat 
P^{I}_{KK^{'}}|\Phi_{\kappa^{'}}\rangle ,\\
&&\mathcal{N}_{\kappa K \kappa^{'}K^{'}}^{I}= \langle \Phi_{\kappa}|\hat P^{I}_{KK^{'}}|\Phi_{\kappa^{'}}\rangle .
 \end{eqnarray*}
 The Hill-Wheeler wave function is given by
\begin{equation}
\psi^{\sigma}_{IM} = \sum_{\kappa,K}~f^{\sigma I}_{\kappa K}~\hat P^{I}_{MK}| 
~ \Phi_{\kappa} \ket.
\label{Anprojaa}
 \end{equation}
where $f^{\sigma I}_{\kappa K}$ are the variational coefficients and index  $^{``}\kappa^{``}$ designates the basis states of Eqns.  (\ref{basis2}) and (\ref{basis3}).
The wave-function is then used to evaluate the electromagnetic 
transition probabilities. 
The reduced electric transition probabilities $B(EL)$ from an initial state 
$( \sigma_i , I_i) $ to a final state $(\sigma_f, I_f)$ are given by \cite{TPSM16}
\begin{equation}
 B(EL,I_i \rightarrow I_f) = {\frac {1} {2 I_i + 1}} 
| \langle \psi^{\sigma_f I_f}|| \hat Q_L || \psi^{\sigma_i I_i} \rangle |^2 ,
  \end{equation}
and the reduced matrix element can be expressed as
\begin{eqnarray*}
 \langle \ \psi^{\sigma_f I_f}|| & \hat Q_L & || \psi^{\sigma_i I_i} \rangle
\nonumber \\ 
&=&\sum_{\kappa_i , \kappa_f, K_i,K_f} {f_{ \kappa_i K_{i}}^{\sigma_i I_i}}~ {f_{ \kappa_f K_{f}}^{\sigma_f I_f}}
 \sum_{M_i , M_f , M} (-)^{I_f - M_f}  \nonumber \\&\times&
\left(
\begin{array}{ccc}
I_f & L & I_i \\
-M_f & M &M_i 
\end{array} \right) 
\nonumber \\
 & & \times \langle \Phi | {\hat{P}^{I_f}}_{K_f M_f} \hat Q_{LM}
\hat{P}^{I_i}_{K_i M_i} | \Phi  \rangle
\nonumber \\
 &=& 2 \sum_{\kappa_i , \kappa_f,K_i,K_f} {f_{ \kappa_i K_{i}}^{\sigma_i I_i}}~ {f_{ \kappa_f K_{f}}^{\sigma_f I_f}}
\nonumber \\
 & & \times \sum_{M^\prime,M^{\prime\prime}} (-)^{I_f-K_f} (2 I_f + 1)^{-1}
\left( 
\begin{array}{ccc}
I_f & L & I_i \\
-K_{f} & M^\prime & M^{\prime\prime}
\end{array} \right)\\
 & & \times \int d\Omega \,D^{I_i}_{M''K_{i}}(\Omega)
\langle\Phi_{\kappa_f}|\hat { O}_{LM'}\hat R(\Omega)|\Phi_{\kappa_i}\rangle
\end{eqnarray*}

 \begin{table}[htp!]
 \LTcapwidth=0.2\textwidth  
\caption {\small{The axial deformation parameter ($\epsilon$) and triaxial
deformation parameter $\epsilon'$ employed in the calculation for
odd-A mass nuclei. The axial deformation $\epsilon$ is taken
from Refs. \cite{WM44,WM45,IH01,WM46,RB04,pm95}. The asterisk ($*$) on $\epsilon$ is  the deformation value for the positive parity bands
in the $^{183}$Au nucleus.  }}\vspace{0.2cm}
\resizebox{1.0\textwidth}{!}
  {
\begin{tabular}{c|ccccccccccccccc}
\hline      & $^{161}$Lu &  $^{163}$Lu   &  $^{165}$Lu  &$^{167}$Lu  & $^{167}$Ta& $^{105}$Pd &$^{127}$Xe & $^{131}$Cs & $^{133}$Ba & $^{133}$La  & $^{135}$Pr & $^{151}$Eu &$^{183}$Au&$^{185}$Au&$^{187}$Au \\
\hline
$\epsilon$   &0.400      &  0.400       &  0.380     & 0.430      & 0.370    & 0.257     & 0.150    & 0.140      & 0.150      & 0.150    & 0.160    & 0.200 & 0.280 &0.270$^*$ &0.220\\
$\epsilon'$  & 0.110     & 0.100        & 0.110      & 0.110      & 0.100    & 0.110     &0.100     & 0.100      & 0.100      &0.110     &0.100     &0.100  &0.110  &0.100    & 0.110\\
 \ $\gamma $ & $15^0$    &  $ 14^0$     &   $16^0$    & $14^0$     & $15^0$   & $23^0$    & $34^0$   &  $36^0$    & $34^0$     & $36^0$    & $32^0$  & $27^0$ & $21^0$& $20^0$  &$27^0$ \\ \hline
\end{tabular}
\label{Tab:Tcr}
}
\end{table}
 %
 
  
\begin{figure}[htbp]
\centering
\includegraphics[width=13cm,height=8cm]{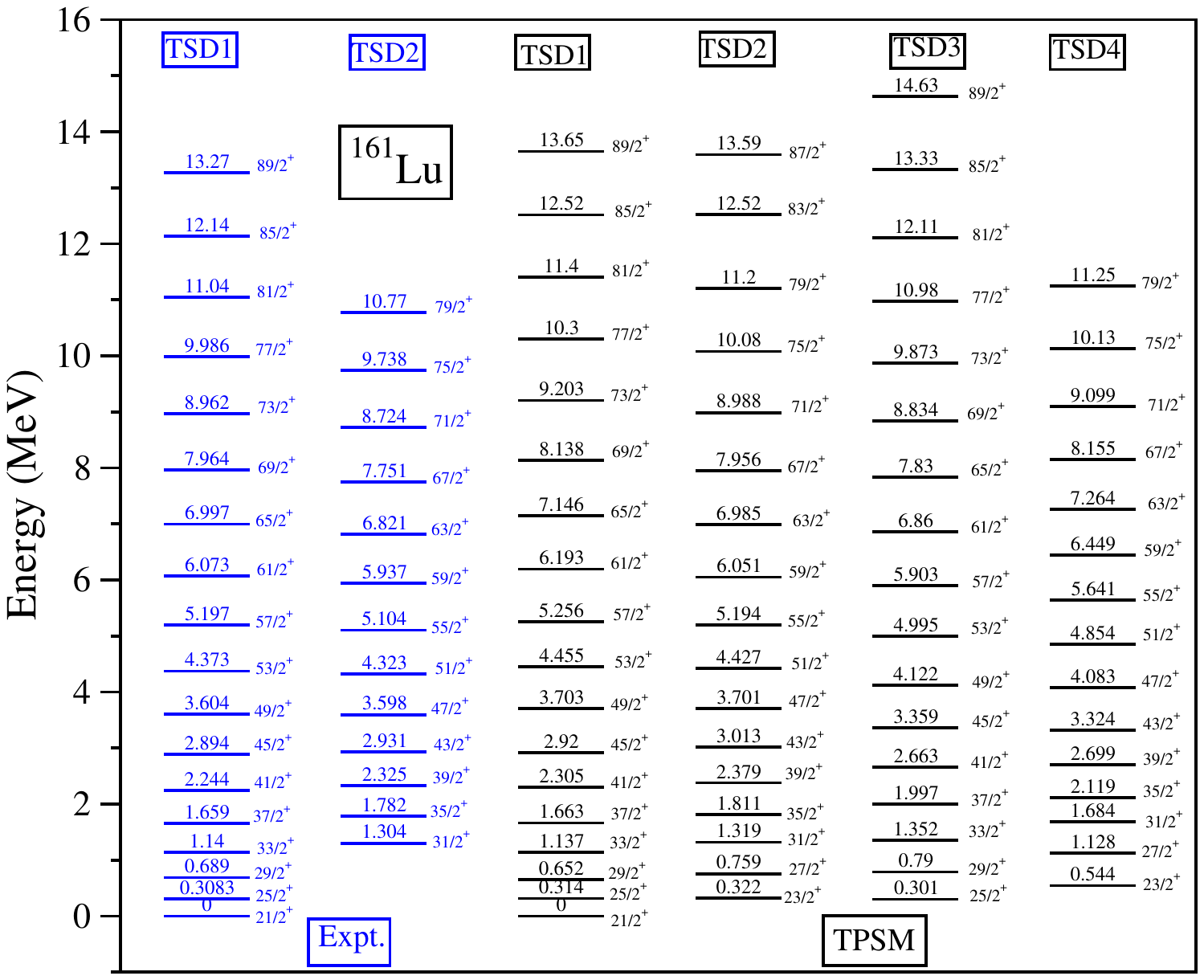}
\caption{TPSM energies for the lowest four bands after configuration
mixing are plotted along
with the available experimental data for  the $^{161}$Lu  isotope Data is
taken from \cite{WM2}.}
\label{161Lu_theoexpt.pdf}   
\end{figure}
\begin{figure}[htbp]
\centering
\includegraphics[width=13cm,height=8cm]{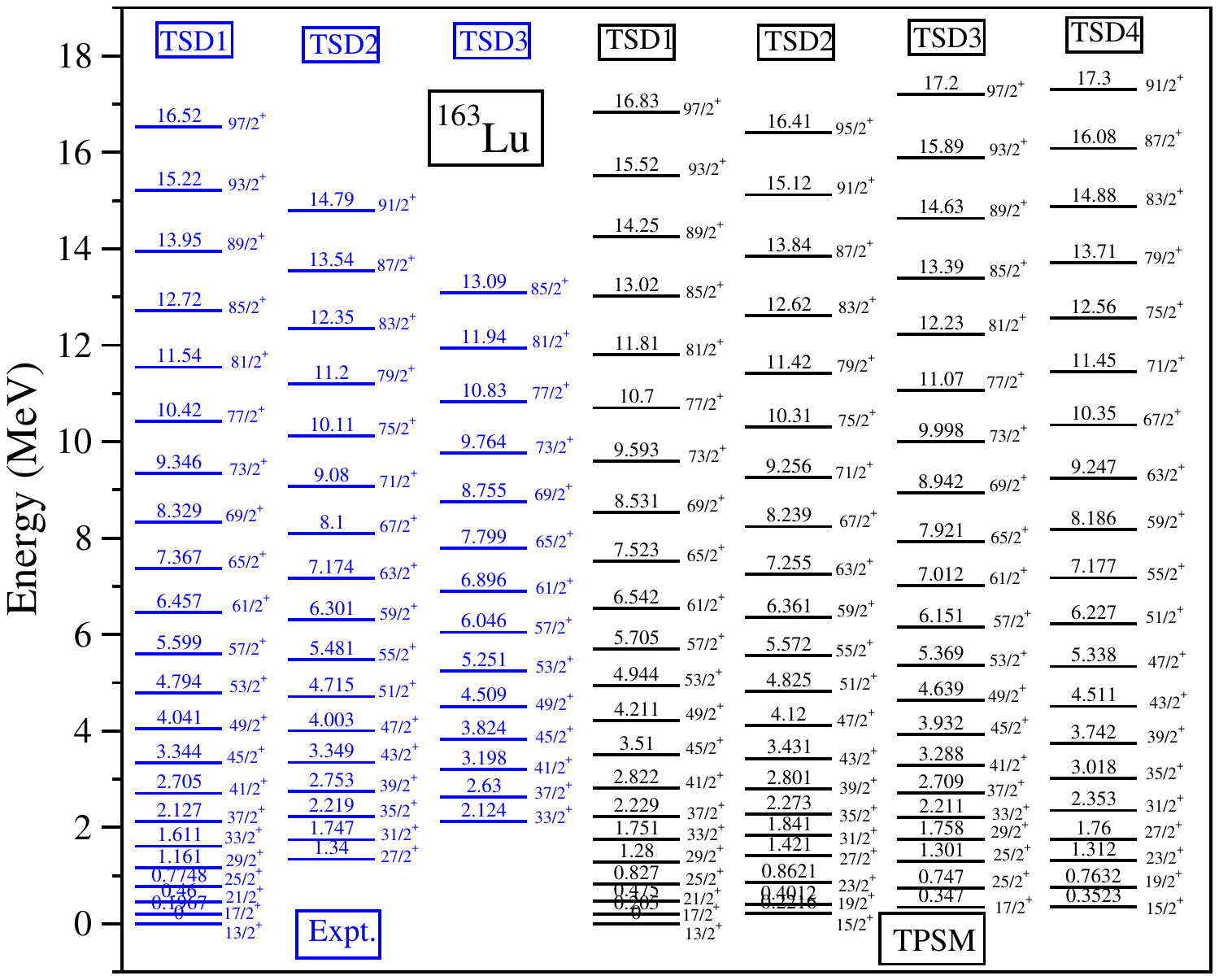}
\caption{TPSM energies for the lowest four bands after configuration
mixing are plotted along
with the available experimental data for the $^{163}$Lu  isotope. Data is
taken from \cite{WM1}.}
\label{163Lu_theoexpt.pdf}   
\end{figure}

\begin{figure}[htbp]
\centering
\includegraphics[width=13cm,height=8cm]{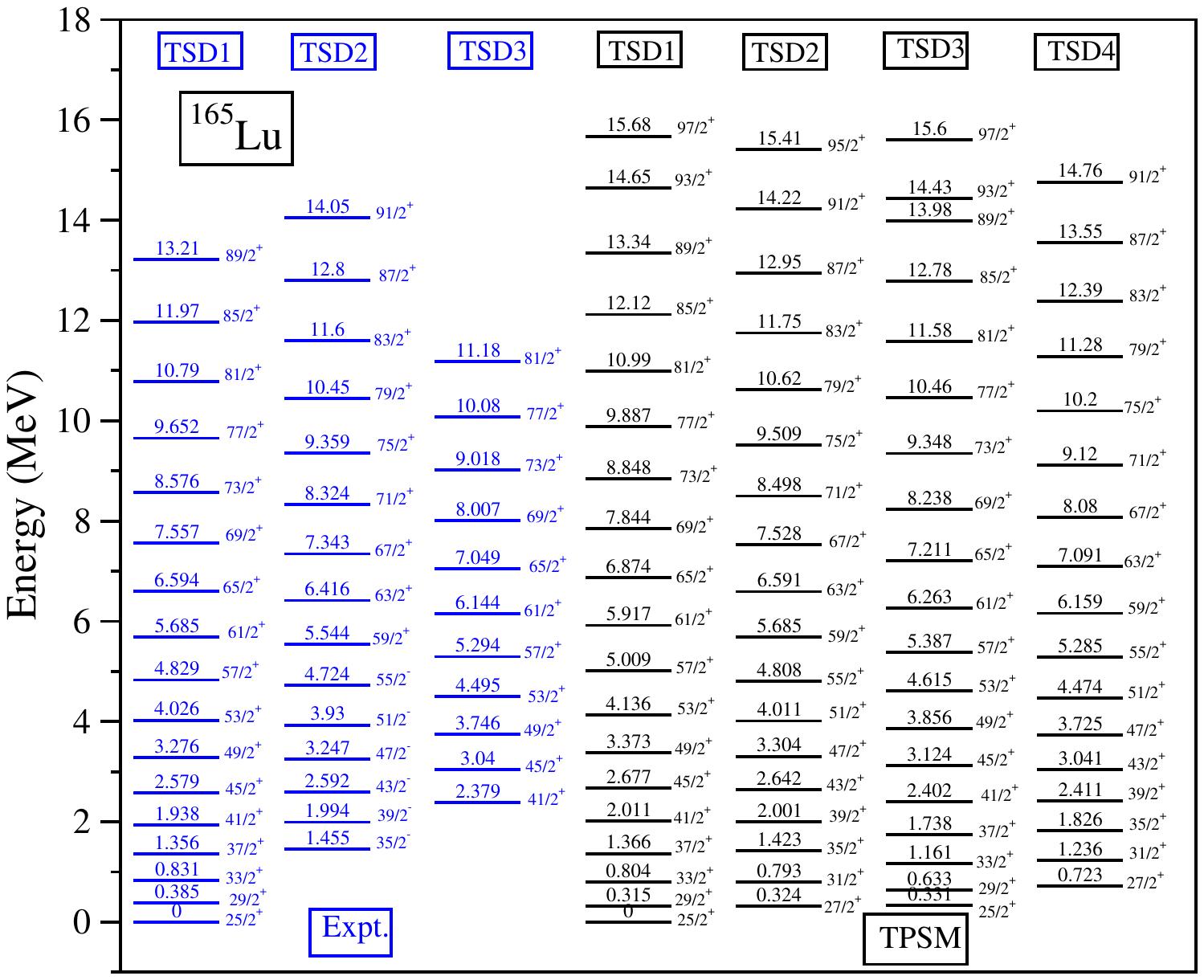}
\caption{TPSM energies for the lowest four bands after configuration
mixing are plotted along
with the available experimental data for the $^{165}$Lu  isotope. Data is
taken from \cite{WM44}.}
\label{165Lu_theoexpt.pdf}   
\end{figure}
\begin{figure}[htbp]
\centering
\includegraphics[width=13cm,height=8cm]{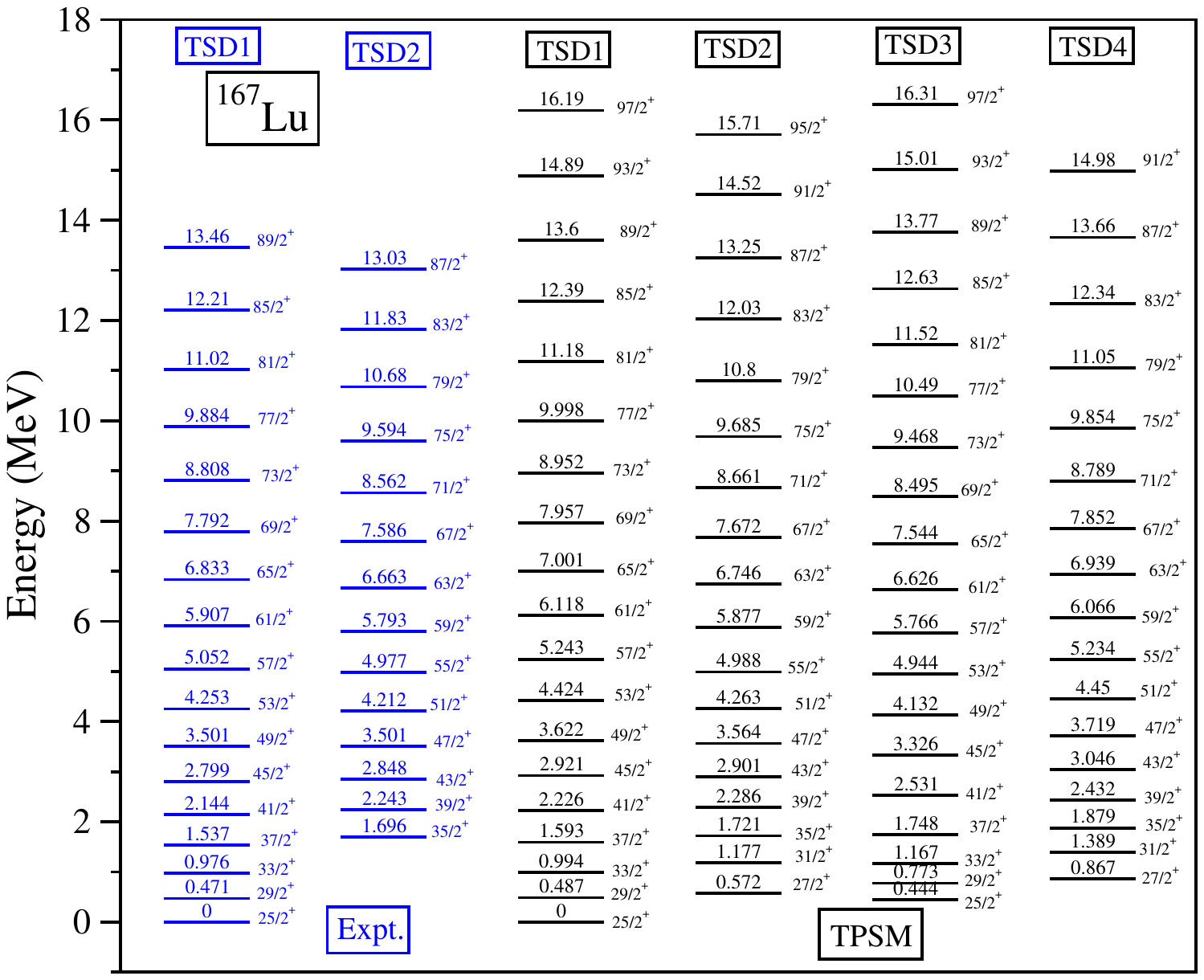}
\caption{TPSM energies for the lowest four bands after configuration
mixing are plotted along
with the available experimental data for  the $^{167}$Lu  isotope. Data is
taken from \cite{HA05}.}
\label{167Lu_theoexpt.pdf}   
\end{figure}

\begin{figure}[htbp]
\centering
\includegraphics[width=13cm,height=8cm]{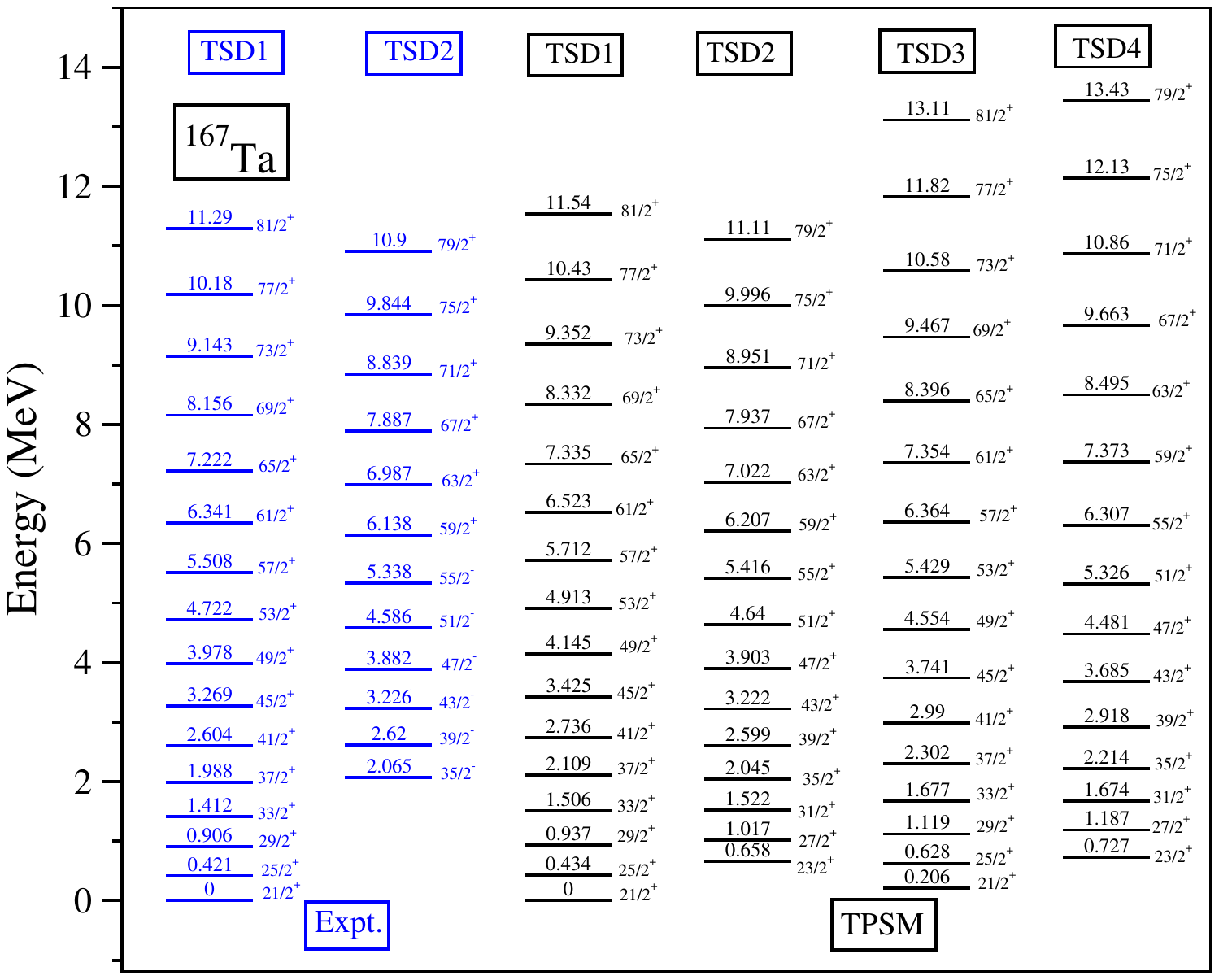}
\caption{ (Color
online) TPSM energies for the lowest four bands after configuration
mixing are plotted along
with the available experimental data for  the $^{167}$Ta  isotope. Data is
taken from \cite{WM46}. }
\label{167Ta_theoexpt.pdf}   
\end{figure}

\section{Results and discussion}

The TPSM calculations for $^{161,163,165,167}$Lu and $^{167}$Ta have been performed with the axial and non-axial deformation
parameters given in Table \ref{Tab:Tcr}. These quantities have been adopted from the earlier studies, in particular, from the ultimate
cranking calculations \cite{WM44,WM45,IH01,WM46,RB04} which predicted TSD shapes for these nuclei.

\begin{figure}[htbp]
\includegraphics[width=6cm,height=6cm]{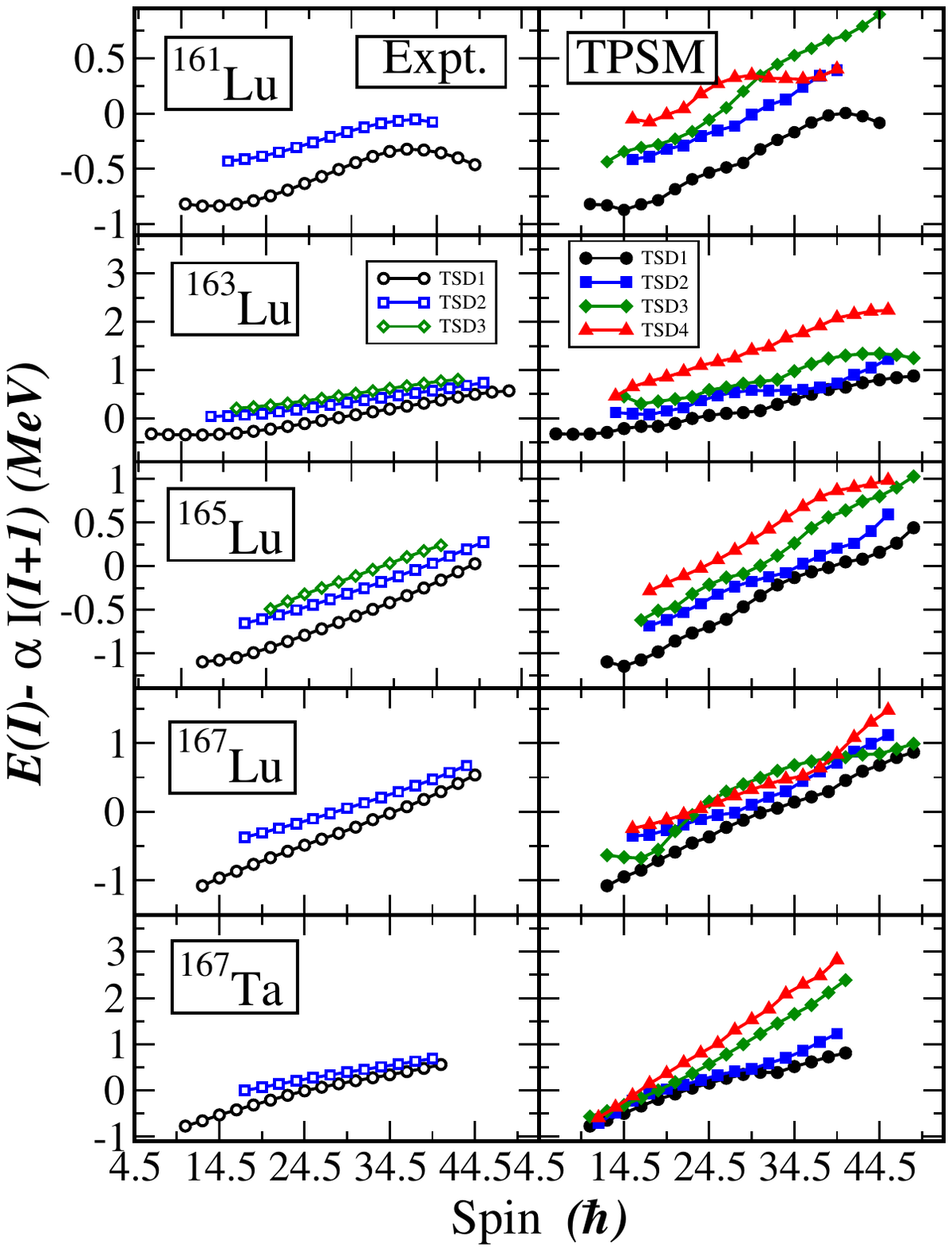}\hspace{0.2cm}
\includegraphics[width=6cm,height=6cm]{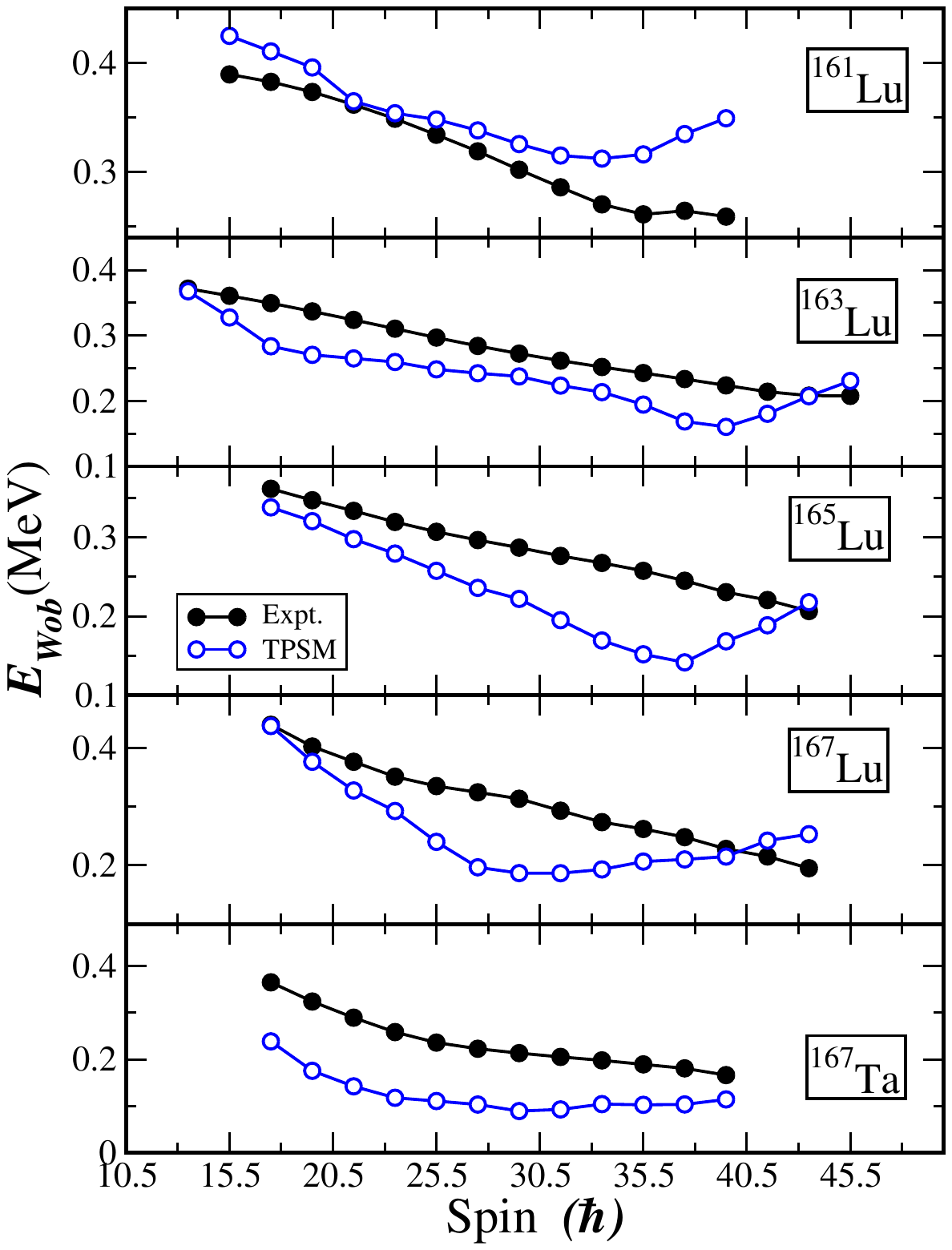}
\caption{ (Color
online) (Left panel) TPSM energies for the lowest four bands after configuration
mixing are plotted along
with the available experimental data for  $^{161,163,165,167}$Lu and $^{167}$Ta  isotopes. The scaling factor $\alpha=32.322A^{-5/3}$. (Right panel) TPSM wobbling energies are compared with  the experimental values obtained from the bands TSD1  and TSD2 for $^{161-167}$Lu and $^{167}$Ta}
\label{energynorm}   
\end{figure}
\begin{figure}[htb]
\includegraphics[width=6cm,height=8cm]{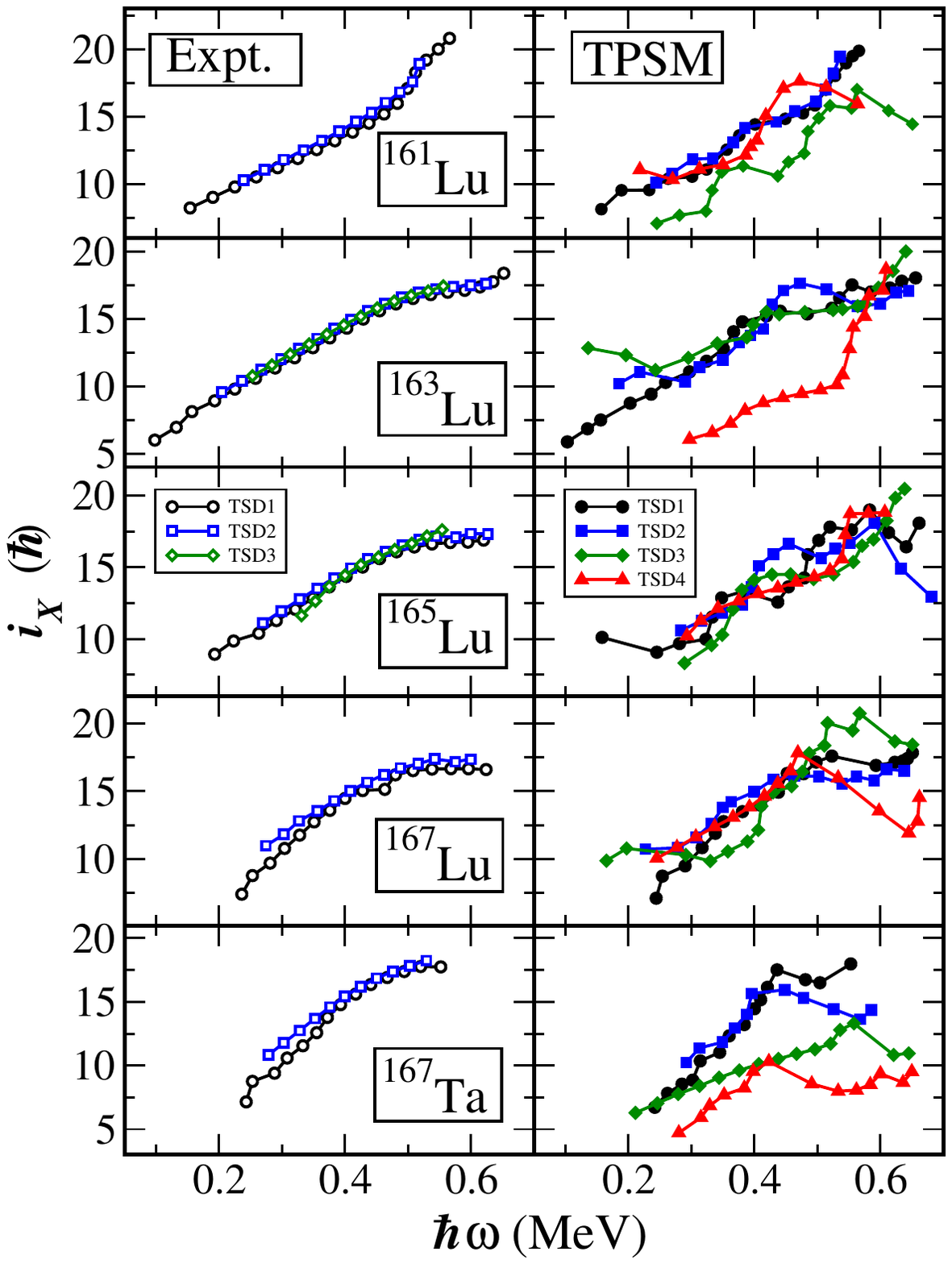}\hspace{0.2cm}
\includegraphics[width=6cm,height=8cm]{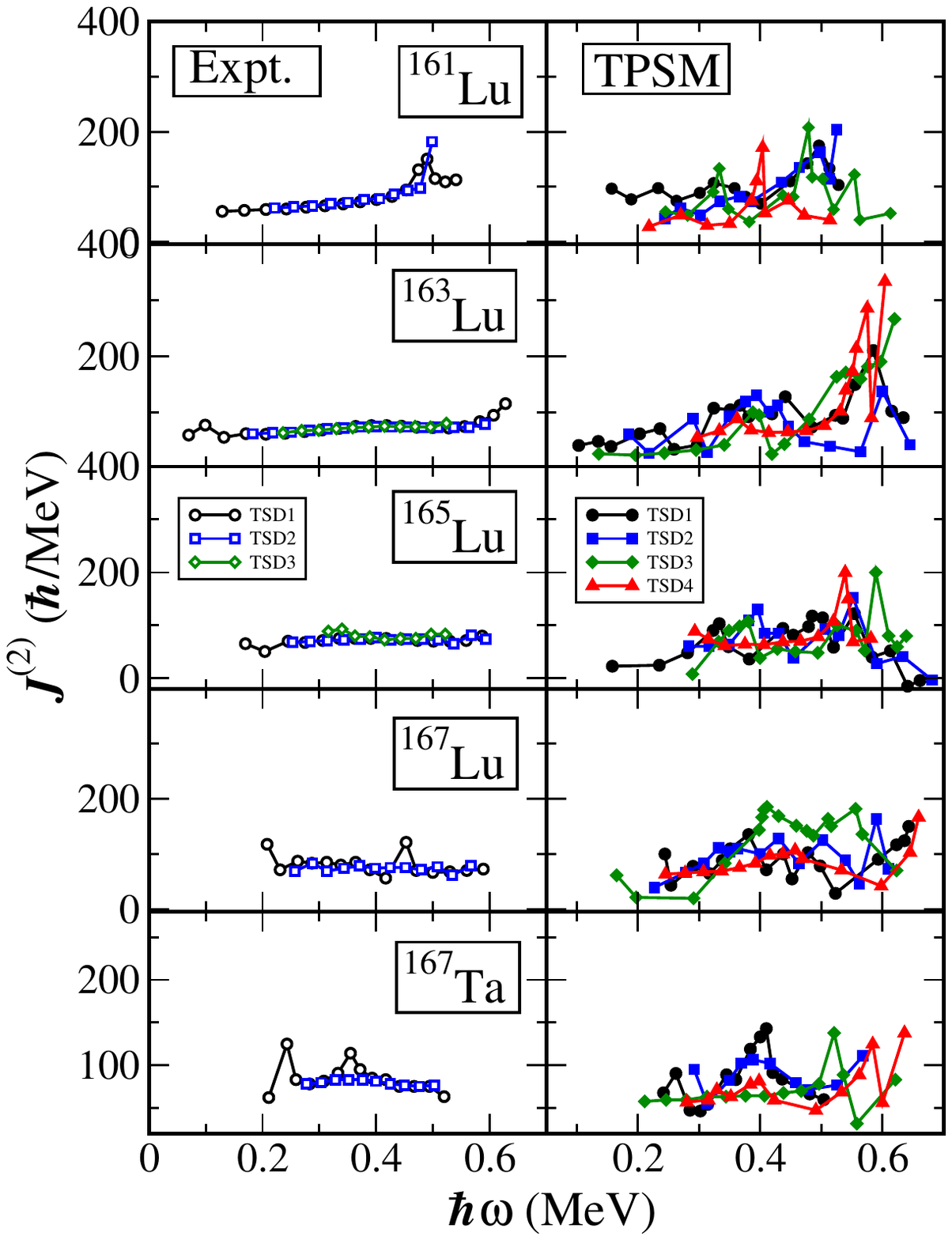}
\caption{(Color
online) (Left panel) Comparison of the aligned angular
 momentum,  $i_x=I_x(\omega)-I_{x,ref}(\omega)$, where $\hbar\omega=\frac{E_{\gamma}}{I_x^i(\omega)-I_x^f(\omega)}$,  $I_x(\omega)= \sqrt{I(I+1)-K^2}$  and $I_{x,ref}(\omega)=\omega(J_0+\omega^{2}J_1)$, obtained from the measured energy levels
 as well as those calculated
 from the TPSM results, for $^{161-167}$Lu and $^{167}$Ta nuclei.
 (Right Panel) Comparison between experimental and calculated dynamic moment of inertia,  ${\mathcal J}^{(2)} =  \frac{4}{E_{\gamma}(I)-E_{\gamma}(I-2)}$, for the TSD1, TSD2, TSD3 and TSD4 for $^{161-167}$Lu and $^{167}$Ta nuclei.  The reference band Harris parameters used are $J_0$=30 and $J_1$=40  \cite{WM44}, obtained from the measured energy levels
 as well as those calculated
from the TPSM results, for $^{161-167}$Lu and $^{167}$Ta nuclei.}
\label{ix.pdf}   
\end{figure}

The TPSM energies obtained after diagonalization
of the shell model Hamiltonian are compared with the available experimental data in Figs.~\ref{161Lu_theoexpt.pdf}, \ref{163Lu_theoexpt.pdf} \ref{165Lu_theoexpt.pdf}, \ref{167Lu_theoexpt.pdf} and \ref{167Ta_theoexpt.pdf}. The calculated
bands are depicted for the lowest two favoured ($\alpha=+1/2$) and two unfavored ($\alpha=-1/2$) signature band
structures. For $^{161}$Lu, two TSD bands have been identified and are compared with the TPSM calculated band structures
in Fig.~\ref{161Lu_theoexpt.pdf}. The experimental bands, labelled as TSD1 and TSD2, have been categorized as $n_{\omega}=0$ and 1 wobbling bands, based
on the similarity of their properties with the corresponding band structures in $^{163}$Lu. It is noted from Fig.~\ref{161Lu_theoexpt.pdf} that
TPSM energies for TSD1 and TSD2 bands are in good agreement with the experimental energies. The deviation of the TPSM energies is small
at low-spin, but with increasing spin the difference  is about 500 keV for the higher observed spin states.
The TPSM energies for $^{163}$Lu are compared with the experimental quantities in the Fig.~\ref{163Lu_theoexpt.pdf} and it
is again evident that
TPSM approach reproduce the known energies quite well with a maximum deviation of about 550 keV at the highest
spin, $I=85/2$. For  $^{163}$Lu, three TSD bands have been identified and have been assigned
as $n_{\omega}$=0,1 and 2. This is the first system where the occurrence of wobbling excitation mode was confirmed for the first time
with the measurement of transition probabilities that will be discussed in detail later. 
\begin{figure}[htb]
\includegraphics[width=6cm,height=8cm]{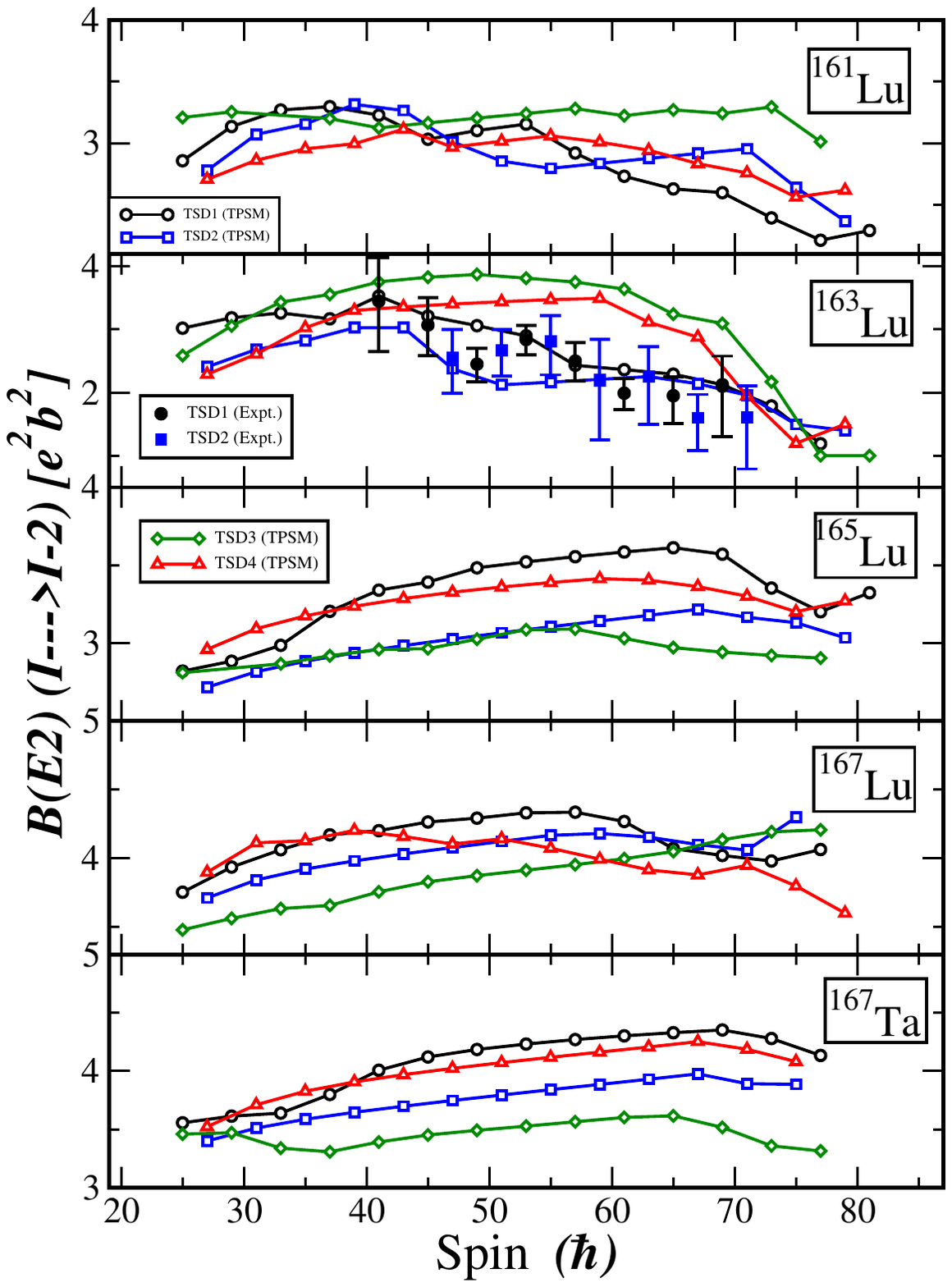}\hspace{0.2cm}
\includegraphics[width=6cm,height=8cm]{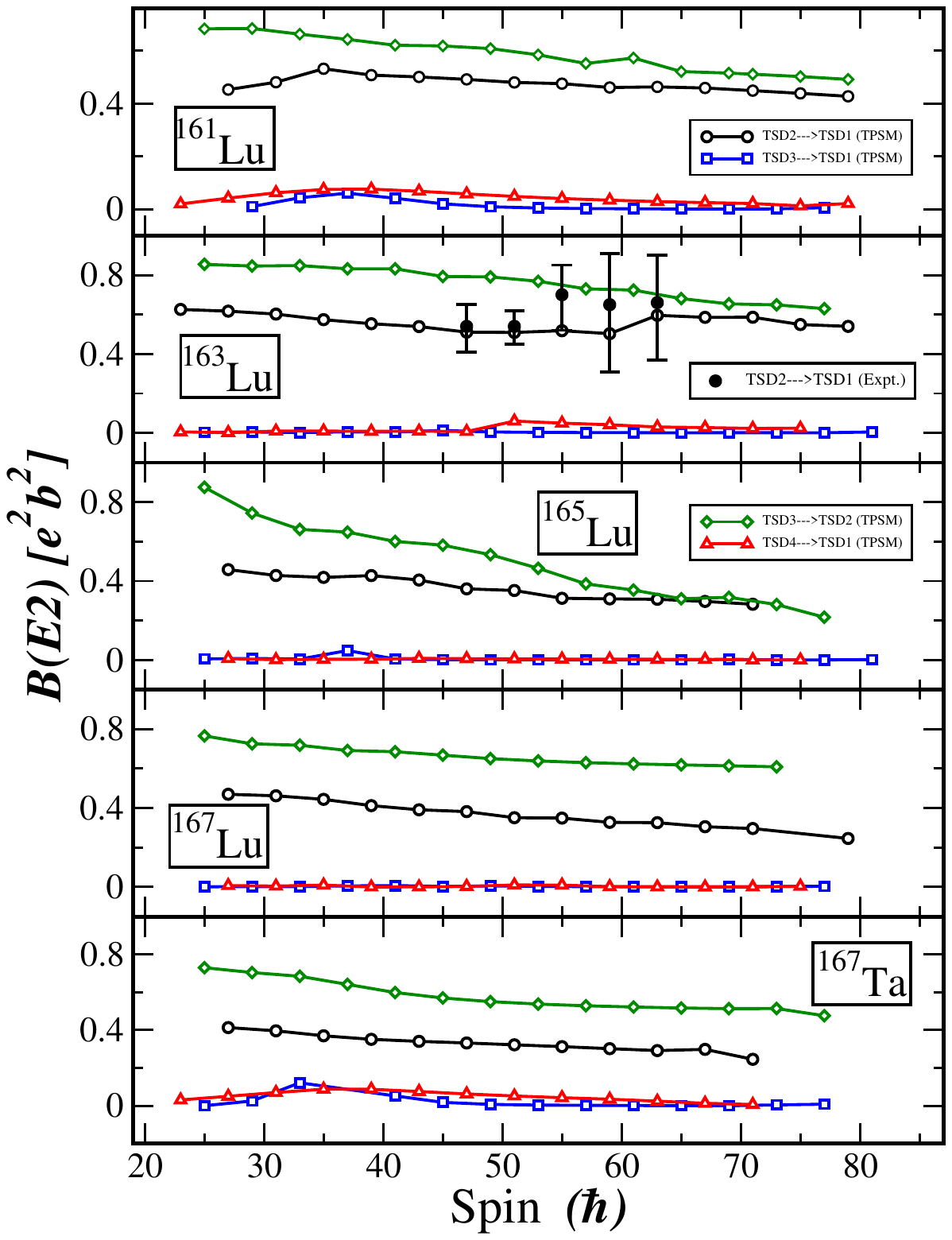}
\caption{(Color
online)  Comparison between experimental and calculated
  $B(\textit{E2})_{in}$ (left panel) and  $B(\textit{E2})_{out}$ (right panel) vs. spin  for the TSD1, TSD2, TSD3 and TSD4 wobbling bands for $^{161-167}$Lu and $^{167}$Ta nuclei.}
\label{BE2_inband.pdf}   
\end{figure}

The calculated TPSM energies for $^{165}$Lu and $^{167}$Lu are compared with the experimental energies in Figs.~\ref{165Lu_theoexpt.pdf} and \ref{167Lu_theoexpt.pdf} . 
Three and two TSD bands have been identified for $^{165}$Lu and $^{167}$Lu, respectively and it is noted again that TPSM
approach reproduces the data quite well. There are differences of about 300 keV for the highest angular momentum states in
both the nuclei. The energies for $^{167}$Ta are compared in Fig.\ref{167Ta_theoexpt.pdf} for the two known TSD bands and it is noted that
TPSM values are in reasonable agreement with the known energies.

To analyze the relative excitation energies of the wobbling bands, for the energies presented in Figs.~~\ref{161Lu_theoexpt.pdf}, \ref{163Lu_theoexpt.pdf} \ref{165Lu_theoexpt.pdf}, \ref{167Lu_theoexpt.pdf}
 and \ref{167Ta_theoexpt.pdf}
a core contribution have been
subtracted, and the resulting energies are presented in Figs.~\ref{energynorm} (left panel). In the wobbling description,
$n_{\omega}=0$ TSD1 band is the result of the rotation of the system about the axis that has the largest
moment of inertia, which is generally the {\it m}-axis. The excited $n_{\omega}=1, 2$ and 3 wobbling bands are obtained when the
angular-momentum from {\it m}-axis is transferred to {\it l}- and {\it s}-axis. The excitation energy of the first wobbling band is about 500 keV and 
the $n_{\omega}=2$ band is less than 500 keV from the $n_{\omega}=1$ band.

\begin{figure}[htb]
\includegraphics[width=6cm,height=8cm]{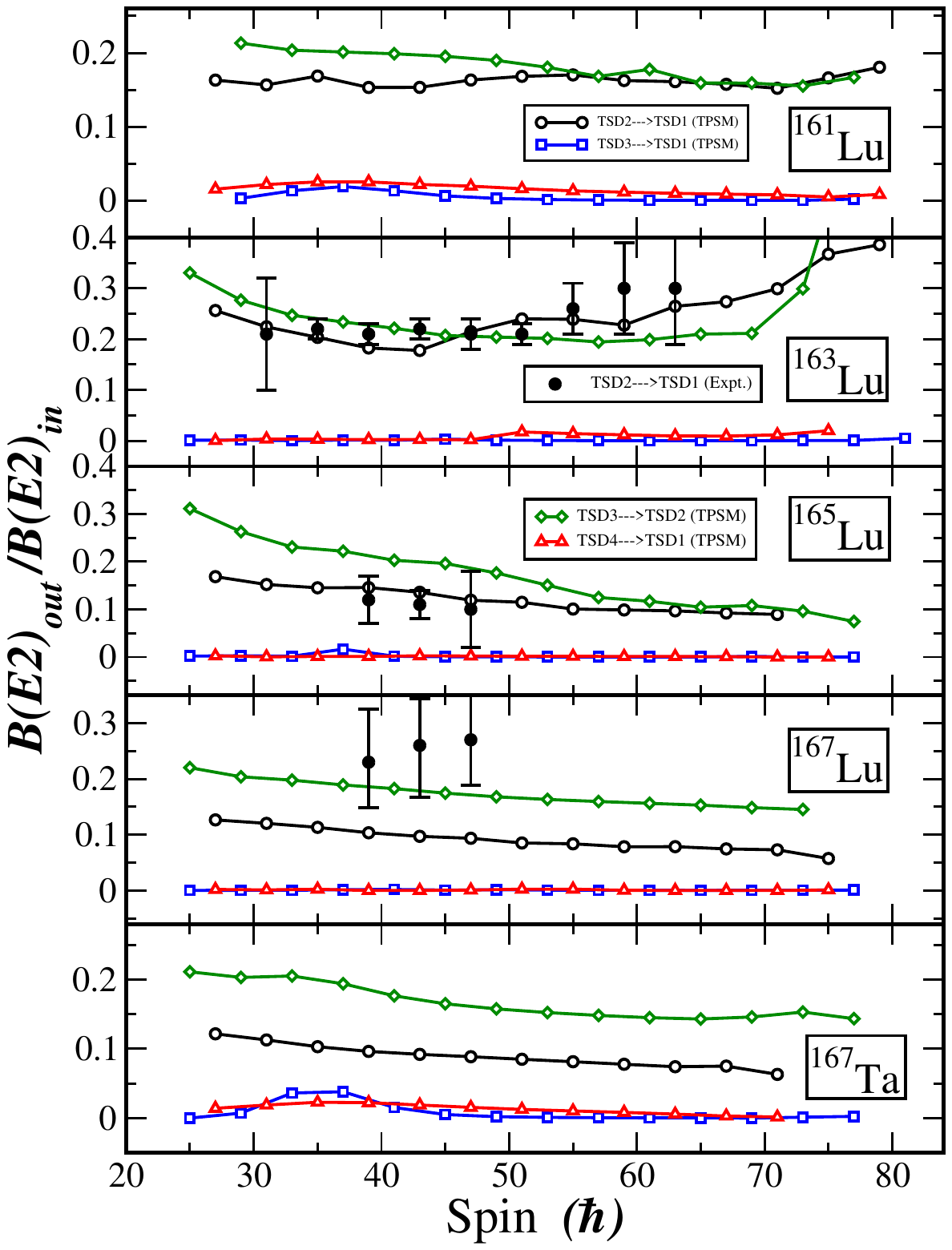}\hspace{0.2cm}
\includegraphics[width=6cm,height=8cm]{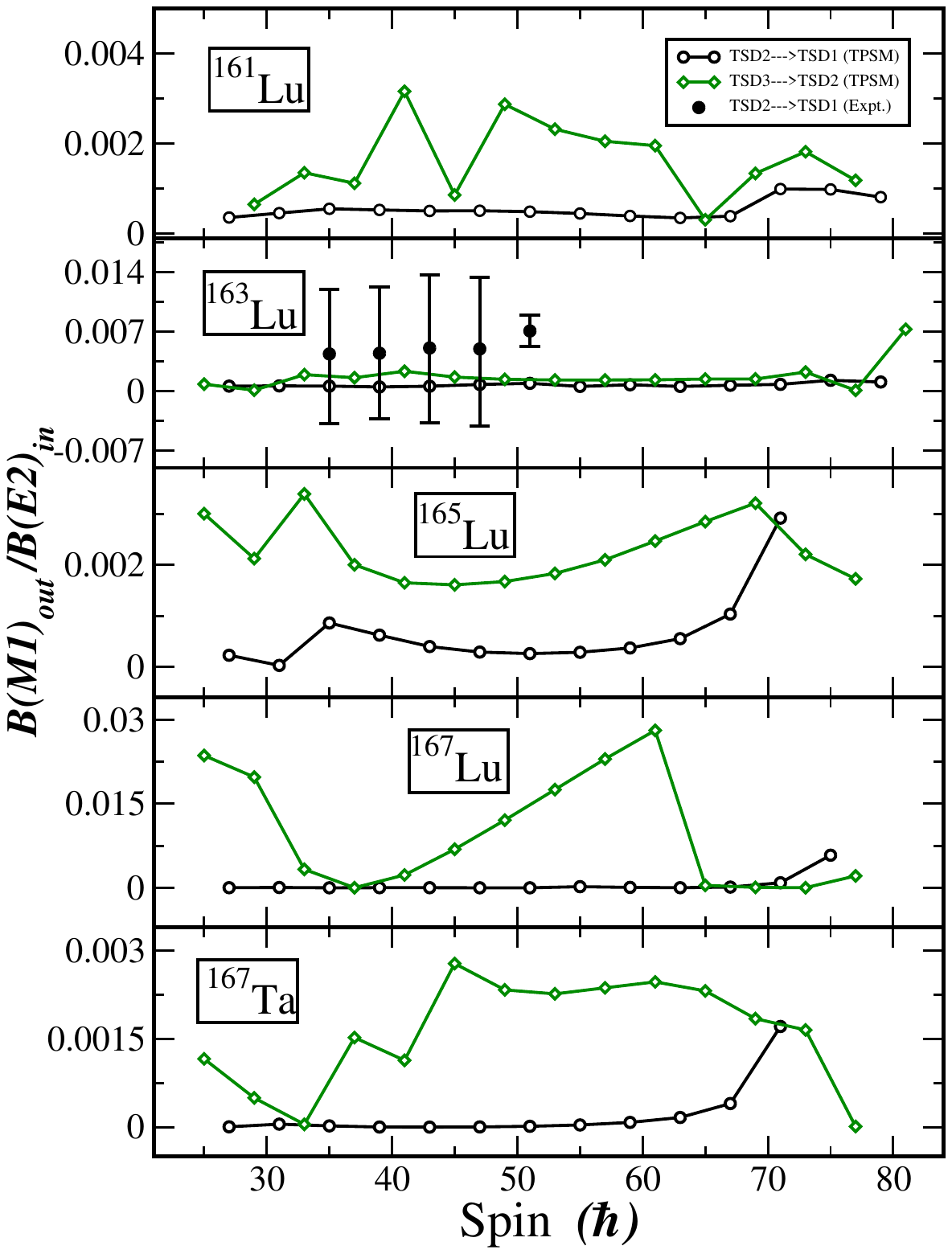}
\caption{(Color
online)  Comparison between experimental and calculated
 $B(\textit{E2})_{out}/B(\textit{E2})_{in}$ (left panel) and $B(\textit{M1})_{out}/B(\textit{E2})_{in}$ (right panel)  vs. spin for the TSD1, TSD2, TSD3 and TSD4  wobbling bands for $^{161-167}$Lu and $^{167}$Ta nuclei.}
\label{BE2O_BE2I_Ratio.pdf}
\end{figure}

For the longitudinal motion, the rotational
axis is along the long-axis and the wobbling frequency, defined as,
\begin{equation}
E_{wob}(I) = E_{n_{\omega}=1}(I) - [E_{n_{\omega}=0}(I+1) + E_{n_{\omega}=0}(I-1)]/2~~~~~~~,
\end{equation}
increases with spin \cite{WM17}. 
For higher values of $n_{\omega}$, the rotational
axis moves away from the {\it m} -axis and will have smaller $K$-values along the {\it m} -axis. 
For the transverse wobbling motion, {\it m} -axis is replaced by {\it s} -axis and the wobbling frequency decreases with spin \cite{WM17}.
The wobbling frequency has been
calculated from the excitation energies spectra and are displayed in Figs.~\ref{energynorm} (right panel) for the
studied Lu- and Ta nuclides. The calculated results are in good agreement
with the experimental values and, in particular, the decreasing tendency of the wobbling energy as a function of angular momentum is
reproduced in all the cases.

As the wobbling bands are based on the same intrinsic configuration, the aligned angular-momentum $i_x$ and moments
of inertia ${J}^{(2)}$ are expected to be identical. These quantities are depicted in Fig.~\ref{ix.pdf} for the
five nuclei. It is noted from the two figures that $i_x$ and ${ J}^{(2)}$  obtained from the measured quantities are almost the
same for all the wobbling bands. These quantities calculated using the TPSM values are also similar for the lowest two
bands TSD1 and TSD2. However, the calculated quantities for the excited bands TSD3 and TSD4 are somewhat different in comparison to the
lowest two bands.

The transition probabilities provide the important information on the nature of the excitation mechanism \cite{BM}. For the
wobbling motion, the probabilities have the characteristic property that transitions from $n_{\omega}$=1 to
$n_{\omega}$=0 are dominated by $B(\textit{E2})$ as compared to $B(\textit{M1})$ in the normal cranking picture for the transitions
from the SP to the yrast band. Further, in the harmonic wobbling limit, the transitions from 
$n_{\omega}$=2 to $n_{\omega}$=1 are predicted to be a factor of two larger as compared to the transitions from $n_{\omega}$=1 to $n_{\omega}$=0,
and direct transition from $n_{\omega}$=2 to $n_{\omega}$=0 are forbidden.

Before discussing the inter-band transitions, we first demonstrate the predictive power of the in-band $B(E2)$ transitions
as these are measured to a good accuracy in some cases. These transitions are shown in Fig.~\ref{BE2_inband.pdf} (left panel)
for the lowest four bands of the five nuclei studied in the present work. The measured values are known for TSD1 and TSD2 bands
of $^{163}$Lu and the TPSM predicted values are noted to be in reasonable agreement with the known transitions. It is evident
from the figure that for $^{161}$Lu and $^{163}$Lu, TSD1 and TSD2 in-band transitions are similar, however, for other bands 
these transitions vary for the four bands. It would be quite interesting to perform the experimental measurements
of the the transition probabilities for other isotopes in order to confirm the varying nature of the predicted values.
\begin{figure}[htbp]
\centering
\includegraphics[width=12cm,height=9cm]{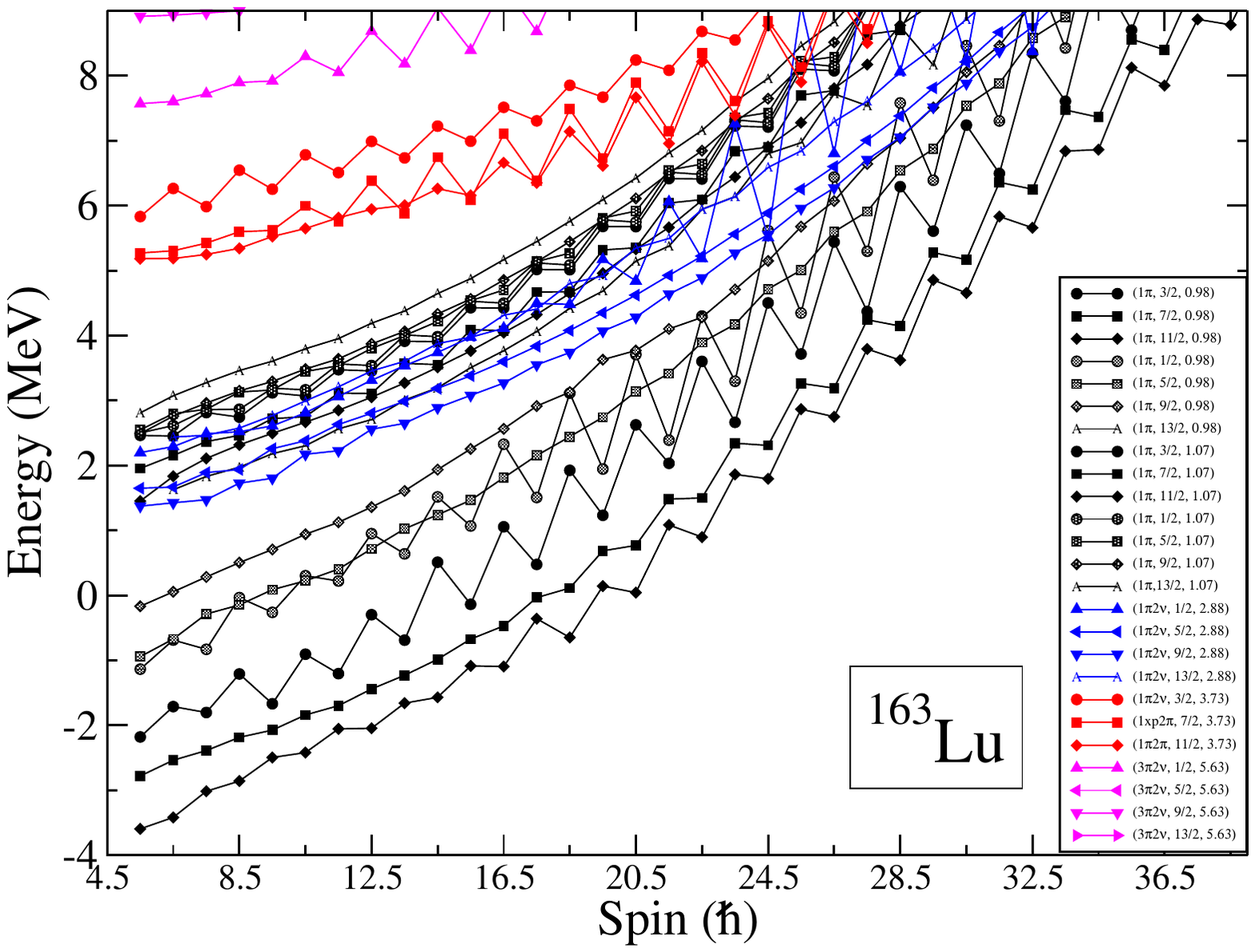}\hspace{0.2cm}
\caption{(Color
  online)   Energies of the projected $K$ -configurations with the short axis chosen as the quantization axis to which ``$K$'' quantum number
          refers. 
     The curves are labelled by three quantities: quasiparticle character, $K$ quantum number and energy of the quasiparticle state.
     For instance, $(1p, 3/2, 0.98)$ designates a one quasiproton state with $K$ = 3/2 having intrinsic energy of 0.98 MeV. }
\label{163Lu_3rdsector_BD_v2.pdf}
\end{figure}

\begin{figure}[htb]
\includegraphics[width=6cm,height=8cm]{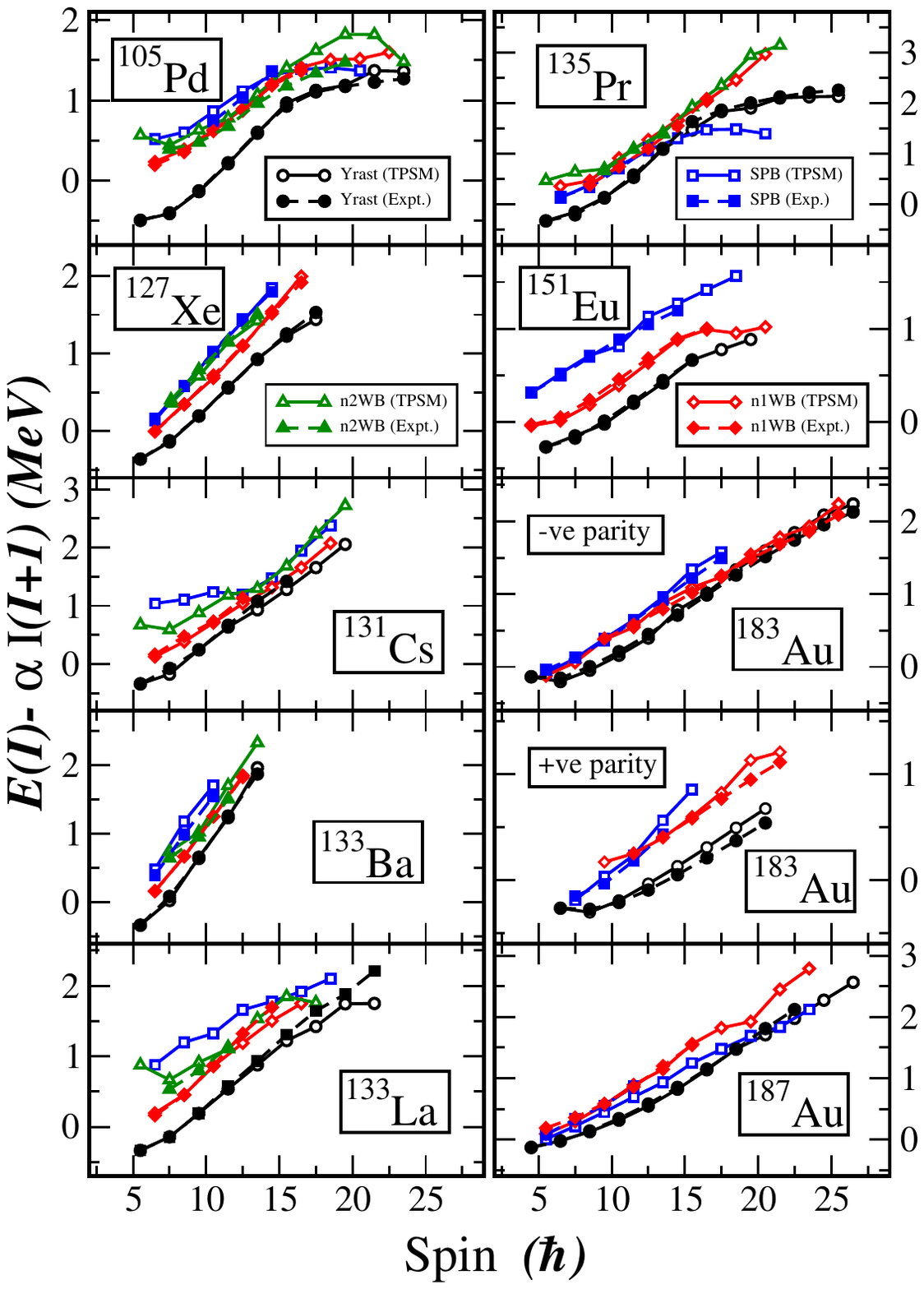}\hspace{0.2cm}
\includegraphics[width=6cm,height=8cm]{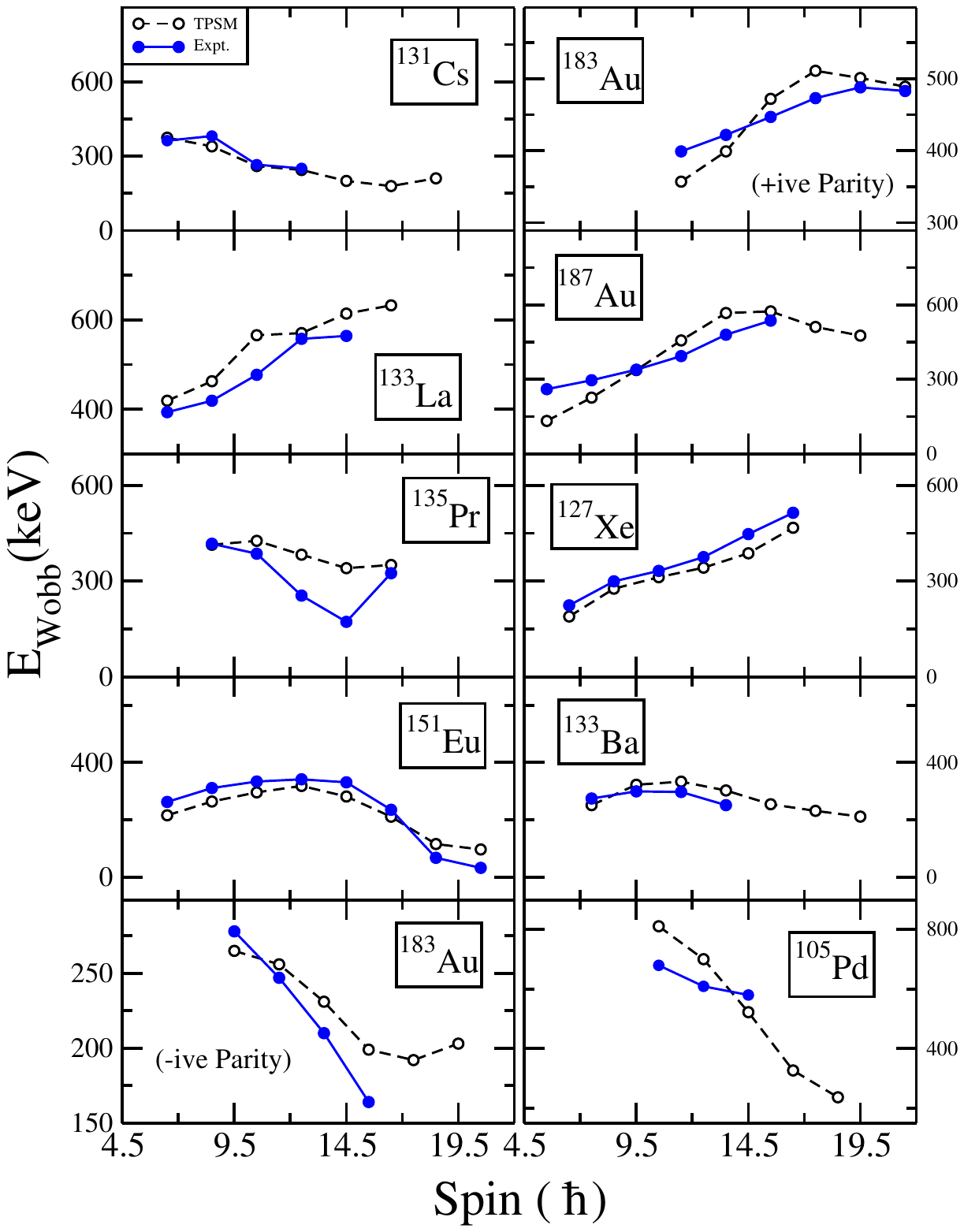}
\caption{(Color
online) (Left panel) TPSM energies for the lowest bands after configuration
mixing are plotted along
with the available experimental data for  $^{105}$Pd, $^{127}$Xe, $^{131}$Cs, $^{133}$Ba, $^{133}$La, $^{135}$Pr, $^{151}$Eu, $^{183}$Au (-ve and +ve parity) and $^{187}$Au nuclei.  The scaling factor $\alpha=32.322A^{-5/3}$.
(Right panel) TPSM wobbling energies are compared with  the experimental values obtained from the bands TW/LW for $^{105}$Pd, $^{127}$Xe, $^{131}$Cs,  $^{133}$La, $^{133}$Ba, $^{135}$Pr, $^{151}$Eu, $^{183}$Au (negative parity), and $^{183}$Au (positive parity). Data is
taken from \cite{WM54,WM47,WM48,WM6,WM12}. }
\label{N1rmenergy}
\end{figure}

The inter-band transitions among the four bands for the five isotopes are depicted in Fig.~\ref{BE2_inband.pdf} (right panel). The
$B(\textit{E2})$ transitions
from TSD3 to TSD2 are clearly enhanced as compared to the transitions from TSD2 to TSD1 and direct transitions from TSD4 to TSD1, and
TSD3 to TSD1 are retarded as expected for the harmonic wobbling mode. For $^{163}$Lu, it is noted that TPSM predicted transitions
for TSD2 to TSD1 are in good agreement with the measured transitions.

The TPSM calculated $B(\textit{E2})_{out}/B(\textit{E2})_{in}$
transition ratios are displayed in Fig.~\ref{BE2O_BE2I_Ratio.pdf} (left panel) and are quite
large as expected for the wobbling bands.
The experimental ratios have been deduced from the DCO and polarization measurements for $^{163}$Lu $^{165}$Lu and $^{167}$Lu and are
also displayed in Fig.~\ref{BE2O_BE2I_Ratio.pdf}. It is noted from the figure that for $^{163}$Lu and $^{165}$Lu, the TPSM
values are in good agreement
with the measured values, however, for $^{167}$Lu the TPSM predicted values deviate for the three known data points. The
$B(M1)_{out}/B(E2)_{in}$ ratios are plotted in Fig.~\ref{BE2O_BE2I_Ratio.pdf}~(right panel) and are seen to be quite small
as compared to the $B(E2)_{out}/B(E2)_{in}$
ratios. These ratios have been measured for $^{163}$Lu and TPSM calculated ratios are in good agreement.

\begin{figure}[htb]
\includegraphics[width=6cm,height=8.5cm]{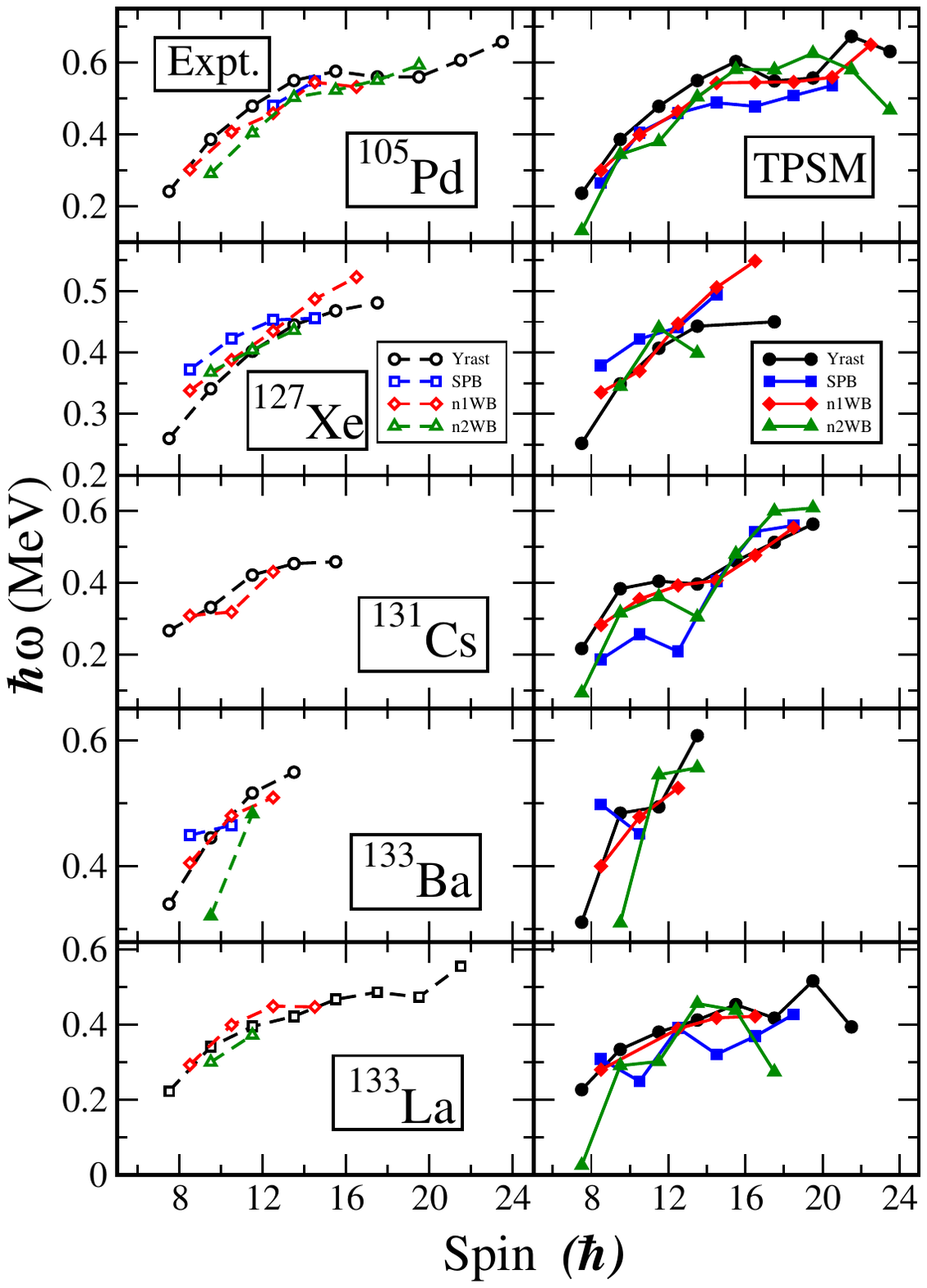}\hspace{0.2cm}
\includegraphics[width=6cm,height=8.5cm]{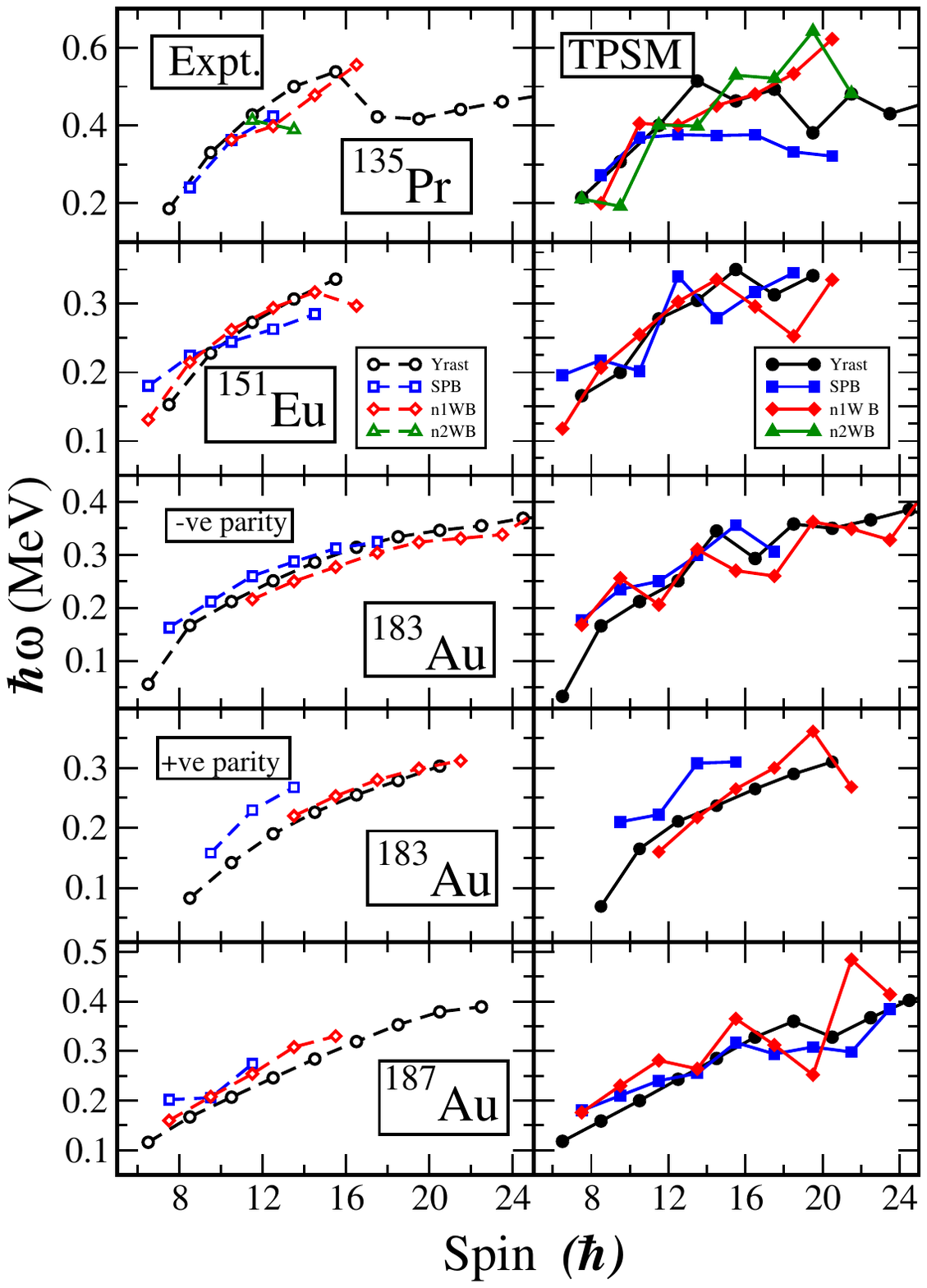}
\caption{ Experimental and calculated values from TPSM for the rotational frequency  as functions of the spin $I$ for the Bands
  $n_{\omega}=0,1$ and 2.}
\label{8fig14}
\end{figure}

To explore further that the above discussed wobbling bands are transverse in character, 
Fig.~\ref{163Lu_3rdsector_BD_v2.pdf} depicts the energies of the bands projected from the quasiparticle configurations before the TPSM
Hamiltonian is diagonalized for $^{163}$Lu as an illustrative example. In the calculations,
the short-axis is chosen as the quantization axis and ``$K$'' denotes the angular momentum projection on this axis. This is
in contrast to the TPSM calculations published so far, where the long-axis is considered as the quantization axis. This choice
simplifies the interpretation since in the TW regime, the odd quasiparticle tends
 to align its angular momentum along the short axis.
 Obviously, such a change of the quantization axis leaves the observables unchanged. Numerically, it is achieved by changing the
 triaxiality parameter $\gamma$ to another equivalent sector that interchanges {\it s}- and {\it l}- axis. For this sector,
 the $\gamma$ value is changed
 to the new value of $(-120^\circ-\gamma)$. It is noted from the figure that the yrast configuration corresponding
 to TSD1 ($\alpha = +1/2$) and TSD2 ($\alpha = -1/2$) band structures have $K$=11/2, which signifies that odd-proton
 is almost aligned towards the {\it s}-axis. The first excited band structures corresponding to TSD3 and TSD4 bands have
 $K$=7/2, indicating that for these band structures the rotational axis has moved away from the principle {\it s}-axis.

\begin{figure}[htb]
  \includegraphics[width=6cm,height=8cm]{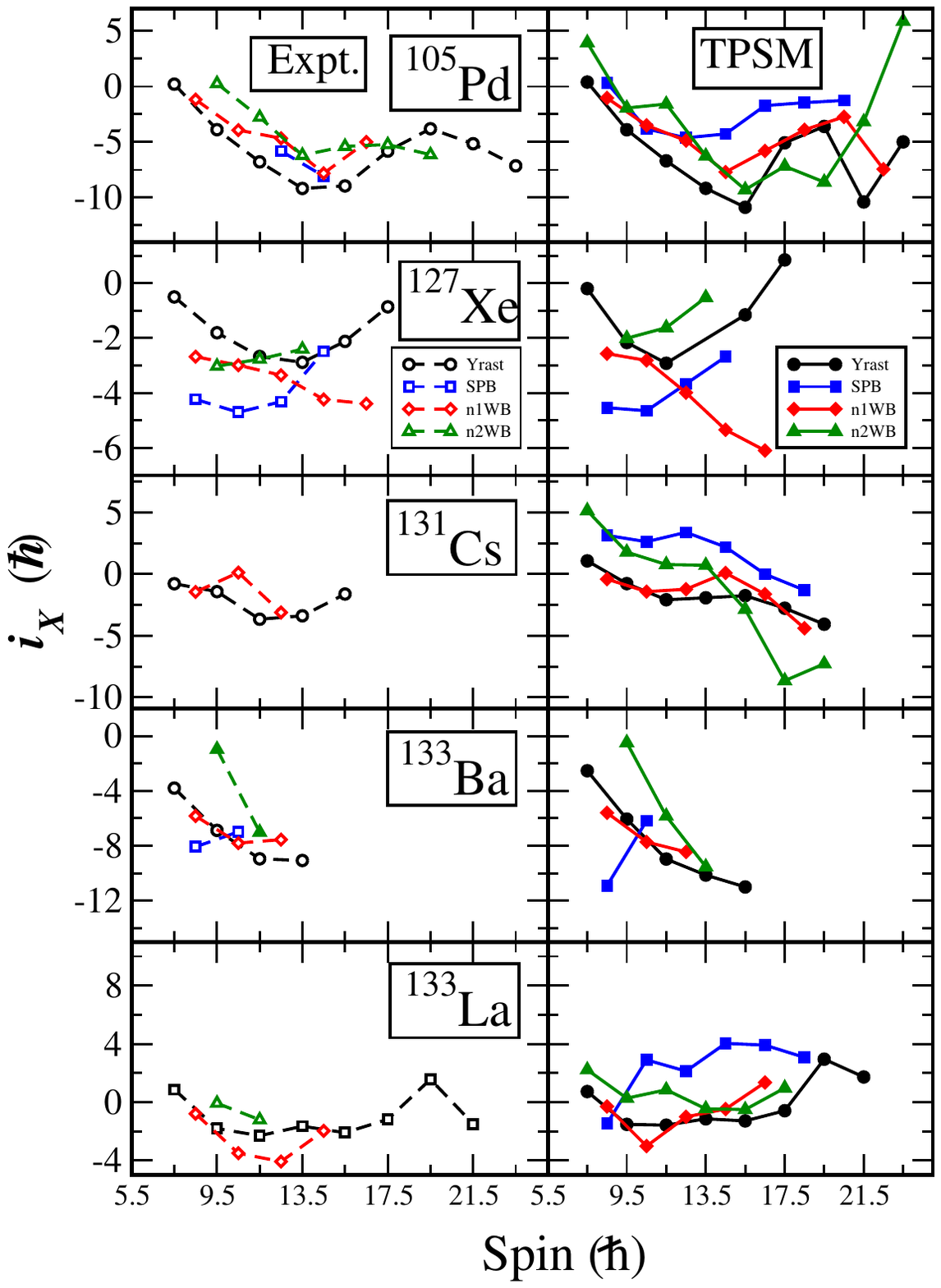}\hspace{0.2cm }
  \includegraphics[width=6cm,height=8cm]{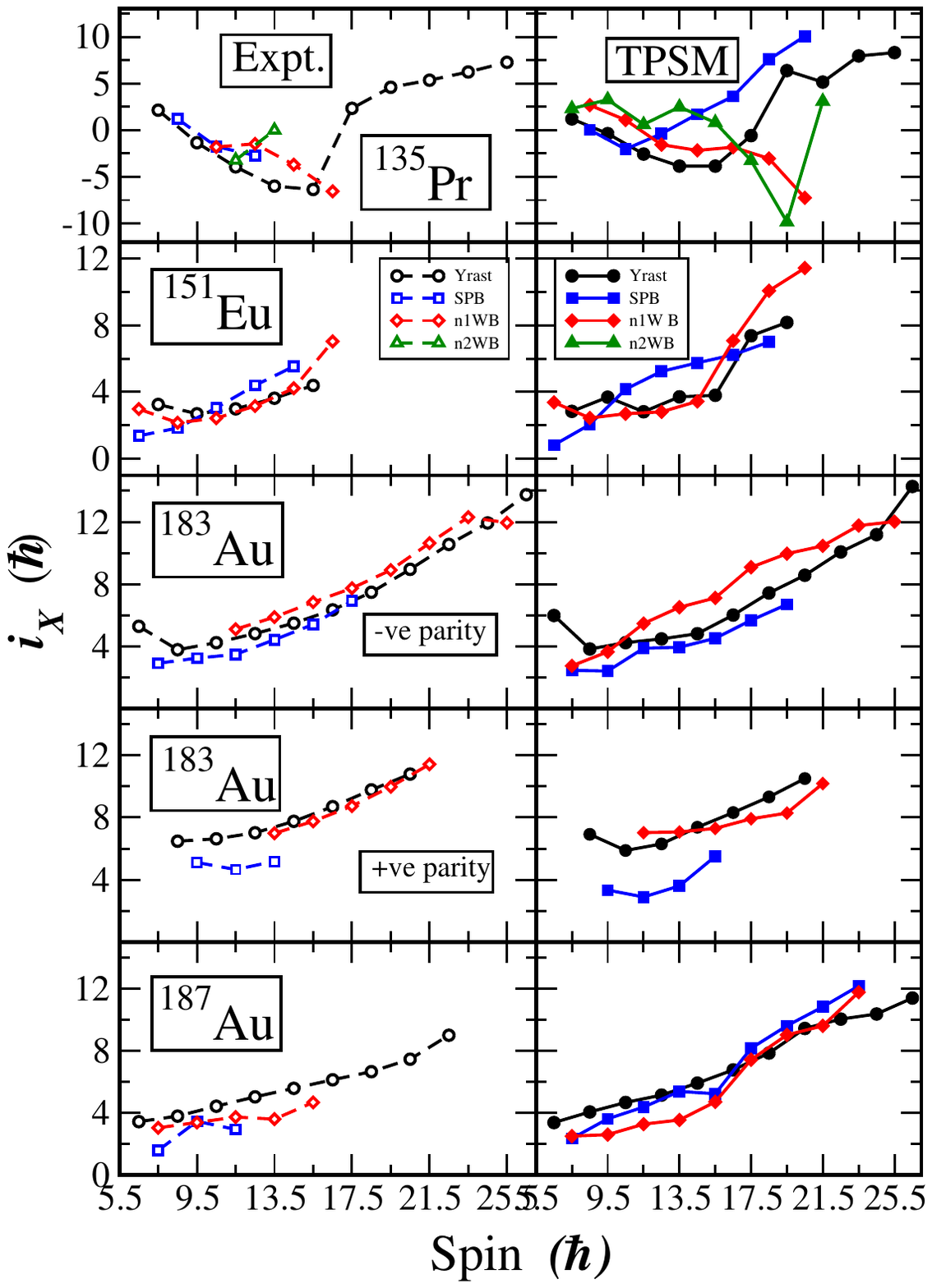}
\caption{ Comparison of the aligned angular
 momentum,  $i_x=I_x(\omega)-I_{x,ref}(\omega)$, where $\hbar\omega=\frac{E_{\gamma}}{I_x^i(\omega)-I_x^f(\omega)}$,  $I_x(\omega)= \sqrt{I(I+1)-K^2}$  and $I_{x,ref}(\omega)=\omega(J_0+\omega^{2}J_1)$, obtained from the measured energy levels
 as well as those calculated
from the TPSM results, for  $^{105}$Pd, $^{127}$Xe, $^{131}$Cs,  $^{133}$La, $^{133}$Ba, $^{135}$Pr, $^{151}$Eu, $^{183}$Au (negative parity), and $^{183}$Au (positive parity) nuclei . The reference band Harris parameters used are  $J_0$=16 and $J_1$=15 \cite{CM96}.
}
\label{Figures/Ix_1_v1}
\end{figure}
%

In the present work, we have also performed the TPSM calculations for ten wobbling bands observed in normal deformed nuclei with the deformation
parameters employed  listed in Table \ref{Tab:Tcr}. On the left panel of Fig.~\ref{N1rmenergy}, the excitation energies for the lowest
bands for all the normal deformed wobbling cases are depicted against the known data and the behaviour of the wobbling frequency as a function of spin is also presented 
in Fig.~\ref{N1rmenergy} on the right panel. The frequency decreases for the cases of $^{131}$Cs, $^{135}$Pr, $^{151}$Eu, $^{183}$Au (negative parity),
$^{133}$Ba and $^{105}$Pd, signifying that these nuclei have transverse wobbling mode. For $^{133}$La,
$^{187}$Au (negative parity) and $^{127}$Xe,
the wobbling frequency increases with spin, which indicates that the collective motion has longitudinal character. For the positive
parity wobbling band in $^{183}$Au, the frequency first increases and then it decreases. It has been predicted for the TW motion that
there is a critical spin up to which the frequency will increase and then after this spin value, the frequency will start
decreasing \cite{WM17}. It is noted that for the negative parity band in $^{183}$Au, the TPSM predicts that the
frequency will increase after I=33/2, and it would be interesting to verify this prediction in future experimental
investigations.
The  derived rotational frequencies  $\hbar\omega$ vs spin (I) are compared with the TPSM  values in Fig.~\ref{8fig14}. The calculations reproduce the experimental data for all the ten wobbling bands observed in normal deformed nuclei.

\begin{figure}[htb]
\includegraphics[width=6cm,height=8cm]{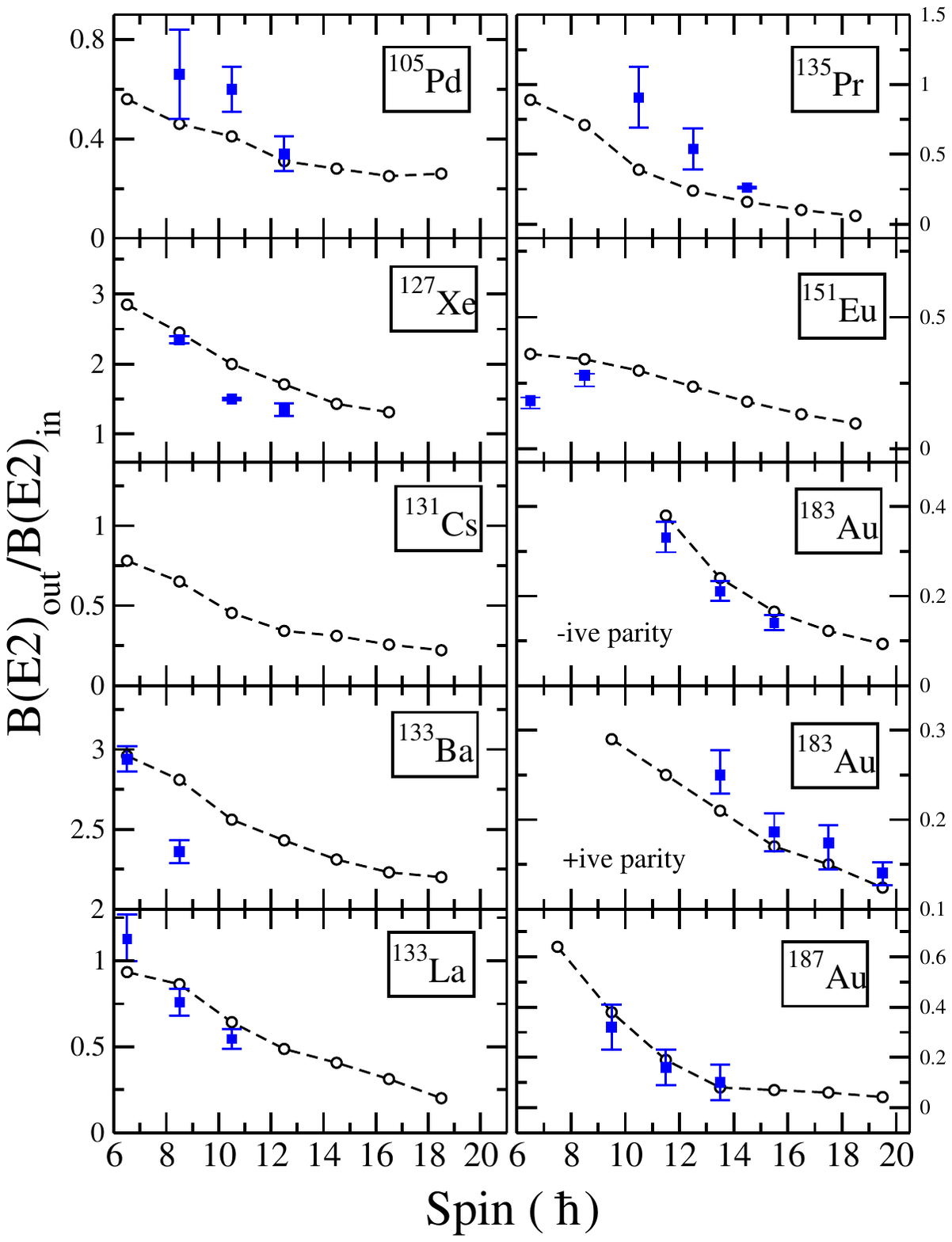}\hspace{0.2cm}
\includegraphics[width=6cm,height=8cm]{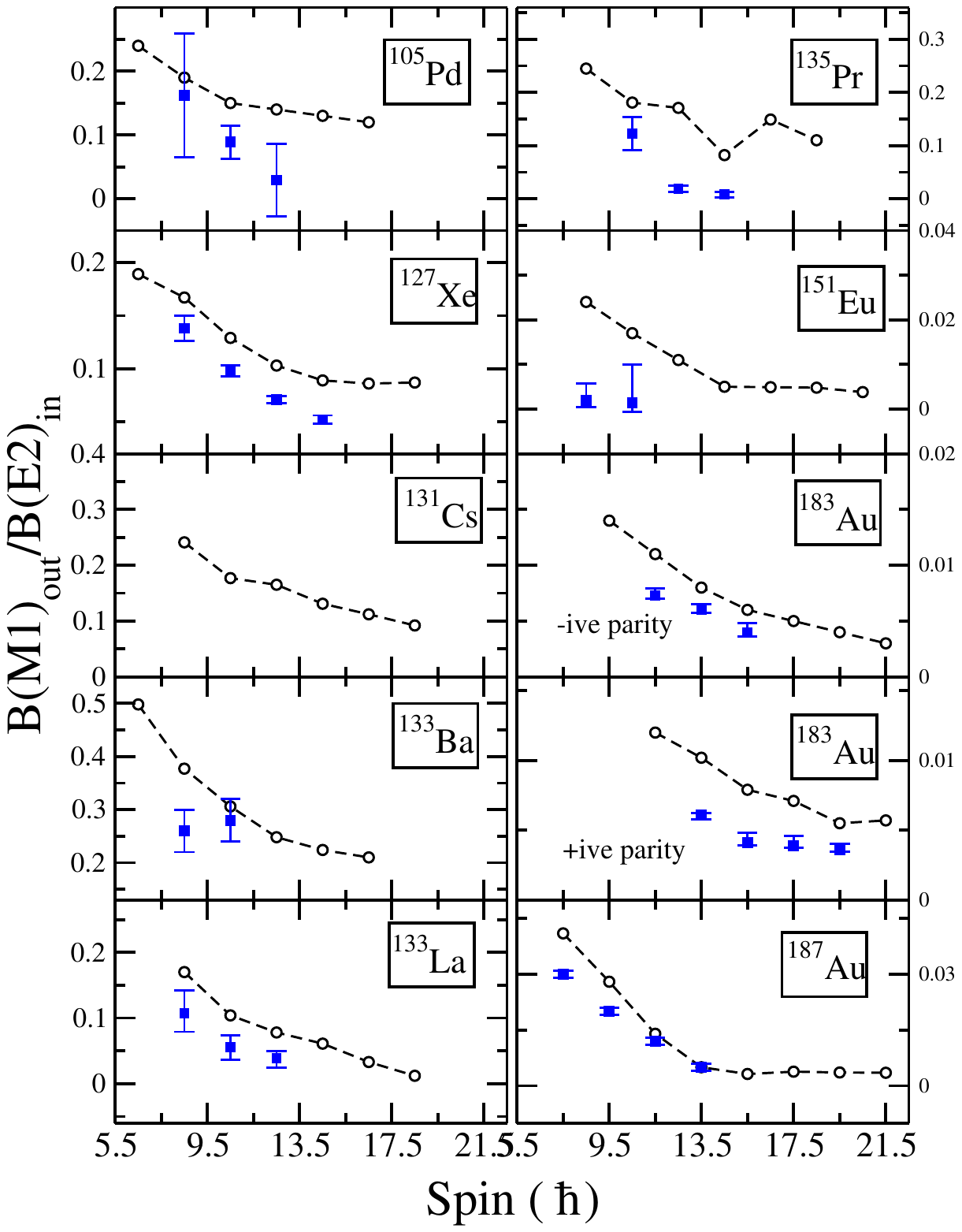}
\caption{Comparison between experimental and calculated
 $B(\textit{E2})_{out}/B(\textit{E2})_{in} $ (left panel) and $B(\textit{M1})_{out}/B(\textit{E2})_{in}$ (right panel) vs. spin for  the bands TW/LW for $^{131}$Cs, $^{133}$La, $^{135}$Pr, $^{151}$Eu, $^{183}$Au ($negative$ parity), $^{183}$Au ($positive$ parity), $^{127}$Xe, $^{133}$Ba and $^{105}$Pd nuclei.}
\label{8fig15}
\end{figure}

To probe further the quasiparticle structures of the observed band
structures, we have analyzed the alignments of the bands as a 
function of the spin and the
results are presented in Fig.~\ref{Figures/Ix_1_v1}. The observed values are in the  reasonable agreement with data.
$B(E2)_{out}/B(E2)_{in}$ transitions for all the 10 nuclides are depicted in Fig.~~\ref{8fig15} (left panel). As already discussed, these transitions are
crucial to establish the wobbling nature of the band structures. It is evident from the ratios that $B(E2)_{out}$ and $B(E2)_{in}$ have
similar values which establishes the wobbling nature of these bands. On the other hand, the ratios $B(M1)_{out}/B(E2)_{in}$ displayed
in Fig.~\ref{8fig15} (right panel) are quite small as expected.

\section{ Summary and future perspectives }

In the present work, we have performed a systematic investigation of the wobbling band structures observed in odd-mass nuclei.
The analysis has been performed for fourteen  nuclides using the triaxial projected shell model approach. This model is now
well established as a method of choice to study the high-spin band structures in deformed and transitional nuclei. The TPSM
approach employs the triaxial basis configurations and is well suited to investigate the properties of triaxial nuclei. It has
been already used to perform a systematic study of the chiral band structures in atomic nuclei, which is a fingerprint of the
triaxial deformation. The wobbling motion is another excitation mode which is only possible for triaxial
shapes. This mode was originally predicted by Bohr and Mottelson \cite{BM} for an even-even system and it was shown that wobbling motion
in the large $I$ limit will give rise to a family of band structures, designated by the harmonic oscillator quantum number, $n_\omega$.
The characteristic feature of the wobbling mode is that transitions $I \rightarrow (I-1)$ from $n_\omega=1$ to $n_\omega=0$ bands are dominated
by $B(E2)$ rather than $B(M1)$ as in the standard cranking model. 

The wobbling mode was first identified in the odd-proton $^{163}$Lu nucleus \cite{WM1} and it was shown that the observed
band structures obey all the characteristic features expected for the wobbling motion. Subsequently, the wobbling bands have
also been observed in other nuclides and in the present work we have studied all these cases using the TPSM approach. It has been demonstrated that
TPSM provides a reasonable description of the observed properties. In particular, the behaviour of the wobbling frequency with
spin is very well reproduced in all the studied cases. The in-band and inter-band transition probabilities have also
been evaluated using the TPSM wavefunctions, and these have been shown to be consistent with the measured data.

In the normal TPSM analysis, long-axis is employed as the quantization axis. In order to analyse the results, we have used the
short-axis as the quantization axis and is achieved by choosing the proper deformation values in another equivalent sector. This change
of the axis simplifies the interpretation for cases where transverse wobbling is expected. For the TW motion, {\it s}-axis is the
rotational axis and the resulting wavefunction will have large components along the {\it s}-axis. This has been shown for $^{163}$Lu
as an illustrative example of TW motion.

Further, the advantage of the TPSM approach is that it can describe the three dimensional wobbling mode and the standard
cranking motion in a unified manner. In the normal cranking mode, SP bands are expected which have dominant $B(M1)$ transition
probabilities, and we have demonstrated that the existence of wobbling and SP band structures
identified in some nuclei can be simultaneously described using the TPSM approach. In the particle-rotor model picture, the wobbling
and SP bands structures have very different geometry. For the case of excited wobbling bands, the rotational angular-momentum
vector is tilted with respect to the principle axis and in the case of SP bands, it is the angular-momentum
of the valence particle that is tilted and not the rotational axis. It is evident  that the TPSM approach assimilates both
these scenarios in a microscopic manner. In a more recent work \cite{AK24}, the TPSM approach has been used to substantiate the existence
of a doublet wobbling excitation mode in $^{105}$Pd.

In future, we intend to investigate the existence of wobbling motion in even-even systems. Several observed band structures in
even-even nuclei have been proposed as candidate wobbling bands. These systems include, $^{130}$Ba, $^{112}$Ru, $^{134}$Ce, $^{104}$Pd, $^{114}$Pd and $^{136,138}$Nd. In a few 
even-even systems, the odd-spin branch of the $\gamma$ band is lower than the even-spin branch \cite{WM17,WM34,JH10,WM15,WM20} in comparison to a
large class of systems where it is opposite. These few
nuclei have been categorized as $\gamma$ rigid and it will be interesting to explore whether these systems have wobbling
characteristics. The common feature is that the first excited band in these few nuclei is the odd-spin branch of the $\gamma$ band
and for the wobbling motion, it is also the odd-spin band having the harmonic oscillator quantum number, $n_{\omega}$=1.
\section{Acknowledgement}
The authors would like to acknowledge Nazira Nazir and S.P. Rouoof for their help in the preparation of some of the figures
presented in the manuscript.


\end{document}